\documentclass[%
 reprint,
showpacs,
 amsmath,amssymb,
 aps,
prd,
notitlepage,
onecolumn
]{revtex4-1}

\usepackage{graphicx}
\usepackage{dcolumn}
\usepackage{bm}

\usepackage{amstext,amsmath,amssymb,amsfonts,bbm,euscript,color,dsfont}
\usepackage[latin1]{inputenc}
\usepackage{epsfig}
\usepackage{hyperref}
\usepackage{amsthm}
\usepackage{subfigure}
\usepackage{color}
\usepackage{multirow}

\usepackage{verbatim} 
\usepackage{graphicx}
\usepackage{latexsym}
\usepackage{array}

\newcommand{\bea}{\begin{eqnarray}}	
\newcommand{\eea}{\end{eqnarray}}
\newcommand{\be}{\begin{equation}}	
\newcommand{\ee}{\end{equation}}
\newcommand{\beq}{\begin{equation}}	
\newcommand{\eeq}{\end{equation}}

\newcommand{\Z}{{\mathbb Z}}
\newcommand{\C}{{\mathbb C}}

\newcommand{\vev}[1]{\left\langle{#1}\right\rangle}

\newcommand{\dd}{{\textrm{d}}}

\def\R{\relax\ifmmode {\mathbb R}  \else${\mathbb R}$\fi}
\def\C{\relax\ifmmode {\mathbb C}  \else${\mathbb C}$\fi}
\def\Z{\relax\ifmmode {\mathbb Z}  \else${\mathbb Z}$\fi}
\def\N{\relax\ifmmode {\mathbb N}  \else${\mathbb N}$\fi}
\def\I{\relax\ifmmode {\mathbb I}  \else${\mathbb I}$\fi}

\begin{document}


\title{The ghost-gluon vertex in the presence of the Gribov horizon}  




\author{B.~W.~Mintz}\email{brunomintz@gmail.com} 

\affiliation{UERJ $-$ Universidade do Estado do Rio de Janeiro,\\
Departamento de F\'isica Te\'orica, Rua S\~ao Francisco Xavier 524,\\
20550-013, Maracan\~a, Rio de Janeiro, Brasil}

\author{L.~F.~Palhares}\email{leticiapalhares@gmail.com}

\affiliation{UERJ $-$ Universidade do Estado do Rio de Janeiro,\\
Departamento de F\'isica Te\'orica, Rua S\~ao Francisco Xavier 524,\\
20550-013, Maracan\~a, Rio de Janeiro, Brasil}

\author{A.~D.~Pereira}\email{a.pereira@thphys.uni-heidelberg.de}

\affiliation{Institut f\"ur Theoretische Physik, Universit\"at Heidelberg,\\
Philosophenweg 12, 69120 Heidelberg, Germany}

\author{S.~P.~Sorella}\email{silvio.sorella@gmail.com}

\affiliation{UERJ $-$ Universidade do Estado do Rio de Janeiro,\\
Departamento de F\'isica Te\'orica, Rua S\~ao Francisco Xavier 524,\\
20550-013, Maracan\~a, Rio de Janeiro, Brasil}

\date{\today}

\begin{abstract}
We consider Yang-Mills theories quantized in the Landau gauge in the presence of the Gribov horizon via the refined Gribov Zwanziger (RGZ) framework.
As the restriction of the gauge path integral to the Gribov region is taken into account, the resulting gauge field propagators display a nontrivial infrared behavior, being very close to the ones observed in lattice gauge field theory simulations. In this work, we explore a higher correlation function in the Refined Gribov-Zwanziger theory: the ghost-gluon interaction vertex, at one-loop level. We show explicit compatibility with kinematical constraints, as required by the Ward identities of the theory, and  obtain analytical expressions in the limit of vanishing gluon momentum. We find that the RGZ results are nontrivial in the infrared regime, being compatible with lattice YM simulations in both SU(2) and SU(3), as well as with solutions from Schwinger-Dyson equations in different truncation schemes, Functional Renormalization Group analysis, and the RG-improved Curci-Ferrari model.
\end{abstract}

\maketitle


\section{\label{sec:Intro}Introduction}

Being two of the most fundamental and difficult problems in 
theoretical physics since many years, the issues of color 
confinement and chiral symmetry breaking in the strong interactions
have been investigated by many methods in Quantum Field Theory (QFT). 
Some of the most successful of them include the Functional 
Renormalization Group 
\cite{Berges:2000ew,Pawlowski:2005xe}, 
the Schwinger-Dyson equations \cite{Bashir:2012fs,Aguilar:2015bud}, 
or effective models as, for example,  the Nambu-Jona-Lasinio model 
\cite{Nambu-JonaLasinio1961,Klevansky:1992qe} and the Quark-Meson 
model \cite{GellMann:1960np}, among others, with their many 
extensions \cite{Fukushima-PNJL-2003,Schaefer:2007pw}. 
From a discretized QFT perspective, 
Monte Carlo simulations of Quantum Chromodynamics on a spacetime 
lattice  have been long considered a major theoretical cornerstone. 
All these approaches share a common trait, the one of being able to 
capture features of the theory which can not be grasped by 
a standard perturbative expansion. 

Another continuum approach derives from a deep observation 
by V. N. Gribov in \cite{Gribov:1977wm} when analyzing 
gauge-fixed Yang-Mills theories. As 
Gribov argued in \cite{Gribov:1977wm}, the Faddeev-Popov 
quantization of gauge theories is not enough to fully 
eliminate gauge copies from the generating functional. More specifically, there exist field configurations for which the  Faddeev-Popov operator $\EuScript{M}(A)$
possesses nontrivial zero modes which give rise to Gribov copies, {\it i.e.} to equivalent field configurations 
which obey the same gauge condition, meaning that  the  counting of the degrees of freedom in 
the functional integral has not  properly done.

In order to face this problem in the Landau gauge, Gribov suggested to 
constrain the domain of integration in the gauge field  not 
to the whole field space, but rather to a closed region, which 
is now called the Gribov region \footnote{The boundary 
of the Gribov region is called the Gribov horizon.}. 
As long as the gauge field configurations in the path integral
are taken from inside the Gribov region, the determinant 
of the Faddeev-Popov operator is nonzero or, in other 
words, the ghost propagator does not display a pole other 
than the one at vanishing momentum. This nonperturbative 
constraint on the ghost propagator is called the no-pole 
condition.

The restriction to the Gribov region can be effectively 
implemented by the introduction of a weight function in 
the action and then taking the integration domain back to all 
gauge configurations. This weight function is known as Zwanziger's 
horizon function and has been originally cast as a nonlocal 
functional of the gauge field, $H(A)$ in \cite{Zwanziger:1989mf}.
In order to express the resulting action as a local 
functional, one has to introduce auxiliary fields, {\it i.e.}:
a pair of bosonic fields,  $(\varphi$, $\bar\varphi)$, as well as a pair of anticommuting ones, $(\omega$, $\bar\omega)$. 

The resulting Gribov-Zwanziger (GZ) action is then local,
renormalizable and effectively constrains the gauge field 
to the interior of the Gribov region.
A further important development of the GZ framework has taken 
place when it was realized that some dimension-two 
condensates, such as the gluon condensate $\vev{A^2}$,  
would be nonvanishing according to the GZ 
effective action and therefore should be considered from the starting 
Lagrangian. This gave birth to what was called the 
Refined-Gribov-Zwanziger (RGZ) action.
For more technical details on the construction of the 
GZ and RGZ actions and some of its consequences we 
refer to 
\cite{Vandersickel:2012tz,Vandersickel:2011zc,Sobreiro:2005ec,
Dudal:2008sp,Dudal:2010tf,Capri:2015ixa,Capri:2015nzw,
Capri:2016aqq} and references therein.

Although the horizon function has been introduced for mainly
theoretical reasons, it may have relevant implications for 
observables. For example, the presence of Gribov's horizon 
has a significant impact on correlation functions of the 
gauge theory. Such correlation functions, on their turn, 
can be used as 
building blocks for the description of observable quantities 
like particle spectra,  via Bethe-Salpeter or Faddeev 
equations, 
or thermodynamical properties of a finite temperature medium. Indeed, the two-point function of the 
RGZ effective theory compares quite well to lattice results 
in different contexts 
as well as to other nonperturbative approaches, like 
the Dyson-Schwinger equations or the Functional 
Renormalization Group. Note that such an agreement has been 
obtained for the RGZ propagator at the tree-level of the 
effective theory.

In this work, we intend to explore a higher correlation 
function, namely the gluon-ghost-antighost triple vertex, 
within the RGZ framework. Starting from the tree-level RGZ 
action, the 
gluon-ghost vertex is the same as in a purely perturbative 
Yang-Mills theory. However, as loop corrections are 
considered, nontrivial propagators, as well as 
vertices containing the auxiliary fields  which, when integrated out, can 
be recast as nonlocal momentum-dependent gluon vertices, 
give rise to contributions to the correlation function 
 containing the nonperturbative Gribov parameter. In 
this sense, the RGZ framework may be able to probe 
 nonperturbative features of the 
gluon-ghost vertex and  might provide some information on the 
infrared behavior of the Yang-Mills coupling. For more results on the 
vertices of the (R)GZ action in the Landau gauge, we refer the reader to \cite{Gracey:2012wf}.

In Sec. \ref{sec:RGZ-action}, we briefly present the recently 
developed BRST-invariant framework of the Gribov-Zwanziger
theory. Next, in Sec. \ref{sec:3-point}, we present our results 
for the ghost-gluon vertex.  For the benefit of the reader,  the technical details of the calculation 
have been collected in Appendix \ref{sec:appendix-1-loop-diagrams}.
Finally, in Sec. \ref{sec:discussion} we compare our results 
to other nonperturbative methods and discuss some of their 
possible implications for future work in Sec. \ref{sec:finalremarks}.

\section{\label{sec:RGZ-action}The BRST-invariant Refined Gribov-Zwanziger action}

In order to establish our notation, let us first write down the 
action of the Refined Gribov-Zwanziger (RGZ) theory in linear 
covariant gauges. It reads
\begin{eqnarray}
S^{\mathrm{loc}}_{\mathrm{RGZ}}&=& S_{\mathrm{FP}}+S_{m}
+S_{\tau}+S_H,
\label{eq:RGZ-local-action}
\end{eqnarray}
where
\begin{eqnarray}
S_{\mathrm{FP}}&=& \int \dd^dx\,\frac14F_{\mu\nu}^aF_{\mu\nu}^a + \int\dd^dx\left(b^a\partial_{\mu}A^{a}_{\mu}-\frac{\alpha}{2}b^a b^a-\bar{c}^a\EuScript{M}^{ab}(A)c^b\right)
 \label{eq:S-FP}
\end{eqnarray}
is the standard Faddeev-Popov action, while $\EuScript{M}^{ab}(A)$ stands for the Faddeev-Popov operator 
\begin{equation}
\EuScript{M}^{ab}(A)(\bullet)=-\delta^{ab}\partial^2(\bullet) +gf^{abc}\partial_{\mu}(A^{c}_{\mu} \bullet).
\label{npbrst6}
\end{equation}

As discussed in \cite{Fiorentini:2016rwx}
it is possible to introduce a mass term 
for the gluon field, given by
\begin{eqnarray}\label{eq:Ah2-mass-term}
 S_m=\int\dd^dx~\frac{m^2}{2}(A^h)_\mu^a(A^h)_\mu^a. 
\end{eqnarray}
Note that the mass term is not directly given in terms of 
the gauge field $A$, but rather as a function of the 
composite gauge-invariant field $A^h$, which is defined as
\cite{Lavelle:1995ty,Capri:2016aqq}
\begin{equation}\label{eq:def-Ah-local}
A^{h}_{\mu}=h^{\dagger}A_{\mu}h+\frac{i}{g}h^{\dagger}\partial_{\mu}h\,,
\end{equation}
with
\begin{equation}\label{eq:def-h}
h=\mathrm{e}^{ig\xi^a T^a}\equiv \mathrm{e}^{ig\xi},
\end{equation}
where $\xi^a$ is the Stueckelberg field discussed in 
\cite{Lavelle:1995ty,Fiorentini:2016rwx,Capri:2015nzw,
Capri:2015ixa,Capri:2016aqq,Capri:2016gut}. One also 
imposes a transversality constraint on $A^h$, so that
\begin{eqnarray}\label{eq:dAh}
\partial_{\mu}A^{h,a}_{\mu}=0\,,
\end{eqnarray} 
as enforced by the Lagrange multiplier field $\tau$ in
\begin{eqnarray}\label{eq:S-tau}
 S_\tau = \int\dd^dx~i\tau^a\partial_\mu (A^{h})_{\mu}^a
 - \int\dd^dx~\bar\eta^a\,\EuScript{M}^{ab}(A^h)\eta^b.
\end{eqnarray}
Note that the fermionic auxiliary fields $\bar\eta$ and 
$\eta$ in (\ref{eq:S-tau}) appear as a consequence of the 
constraint $\delta(\partial_\mu\,A^h_\mu)$ in the path 
integral expression for the partition function 
\footnote{We thank M. Tissier, N. Wschebor and U. Reinosa 
for having called our attention to this important point.}.

We stress that the constraint (\ref{eq:dAh}) is crucial for 
the renormalizability of the action 
(\ref{eq:RGZ-local-action}), since otherwise a 
propagator like $\langle\xi(p)\xi(-p)\rangle$, for example,
would be ill-defined, leading to non power-counting 
divergences that spoil the renormalizability of the theory
\cite{Ferrari:2004pd}. This is the case in the usual 
formulation of Stueckelberg-like theories. However, in the 
presence of the constraint (\ref{eq:dAh}), all propagators 
are well-behaved and the theory can be shown to be
renormalizable \cite{Fiorentini:2016rwx,Capri:2017bfd}.

Within the RGZ framework, the origin of such a mass term 
is motivated by the fact that, in the presence of the Gribov 
horizon, the theory is unstable with respect to 
the formation of some condensates of operators of mass 
dimension $d=2$. In particular, one can show that 
$\left\langle A^2\right\rangle\not=0$, so that the parameter
$m^2$ can even be interpreted as a Lagrange multiplier that 
ensures that the gluon condensate is nonvanishing in the 
deep infrared 
\cite{Dudal:2007cw,Dudal:2008sp,Gracey:2010cg,Dudal:2011gd,Vandersickel:2011zc}.

The Zwanziger horizon term, in its local form, is given by \footnote{One should note that $\left[\EuScript{M}(A^h)+\mu^2\right]^{ab}$ is a shorthand notation for $\EuScript{M}^{ab}(A^h)+\mu^2\delta^{ab}$.}
\begin{eqnarray}\label{eq:Horizon-local}
 S_H=\int\dd^dx\left(\bar{\varphi}^{ac}_{\mu}\left[\EuScript{M}(A^h)+\mu^2\right]^{ab}\varphi^{bc}_{\mu}-
 \bar{\omega}^{ac}_{\mu}\left[\EuScript{M}(A^h)+\mu^2\right]^{ab}\omega^{bc}_{\mu}+g\gamma^2 f^{abc}(A^{h})_{\mu}^a(\varphi^{bc}_{\mu}-
 \bar{\varphi}^{bc}_{\mu})\right)\,.
\end{eqnarray}

There are some reasons to consider the composite field 
$A^h$ instead of the gauge field $A$ itself in both the 
mass term (\ref{eq:Ah2-mass-term}) and in the local horizon
term (\ref{eq:Horizon-local}). It is crucial to 
recall from \cite{Lavelle:1995ty,Capri:2016aqq}
that $A^h$ defined in (\ref{eq:def-Ah-local}) actually 
corresponds to a local version of the gauge configuration 
that minimizes the functional
\begin{eqnarray}
 f_A[u]\equiv{\rm Tr}\int d^4x\,A_\mu^u(x) A_\mu^u(x)
\end{eqnarray}
along the gauge orbit, parametrized by the gauge transformation 
$u$. Being a minimum of $f_A$ along the gauge orbit, the 
gauge variation of $A^h$ is zero, and so is its BRST variation: $ sA^h=0$, where $s$ is the nilpotent BRST operator. This immediately 
implies the BRST invariance of the mass term $S_m$. Analogously, taking into account that the BRST  variations of the auxiliary 
fields are all zero, the horizon term $S_H$ turns out to be BRST invariant as well \footnote{For the complete set of 
BRST transformations of the fields in the action 
(\ref{eq:RGZ-local-action})  we refer to 
\cite{Capri:2017bfd}.}. 

Finally, we stress that, for practical loop calculations, one 
must expand the $h$ field (\ref{eq:def-h}) in powers of the 
Stueckelberg field $\xi$ up to the desired order. However, 
in the present work, we shall work  in the Landau gauge, $\alpha=0$, in which the $\xi$ field decouples completely 
and no internal lines with $\xi$ propagators are present, see 
\cite{Fiorentini:2016rwx,Capri:2017bfd}. 

Now that the action has been established, let us proceed to the 
explicit calculation of the ghost-gluon vertex in the one-loop 
approximation in the Landau gauge, $\alpha=0$.

\section{The three-point ghost-gluon correlation function}
\label{sec:3-point} 

From the action (\ref{eq:RGZ-local-action}), one can derive 
the Feynman rules of the theory. The rules which are 
relevant to the calculation of the ghost-gluon vertex at one-loop are listed in Appendix \ref{sec:appendix-feynman-rules}.
They allow us to calculate the connected correlation function
\begin{eqnarray}\label{eq:cbcA-def}
 \vev{A_\mu^a(k)\,\bar c^b(p)\,c^c(q)}_{q=-p-k} = \left.\frac{\delta^3Z_c}{\delta (J_{\bar A})_\mu^a(k)\delta J_{\bar c}^b(p)\delta J_{c}^c(q)}\right|_{q=-p-k}
\end{eqnarray}
at one-loop order, where $Z_c$ is the generator of connected 
correlation functions and $J_i$ 
($i=\bar c,c,A$) are external sources linearly coupled to the fields $i$. As usual, the sources are taken to zero at the end of the calculation.


\begin{figure}\label{fig:feyndiags}
\includegraphics[height=8cm]{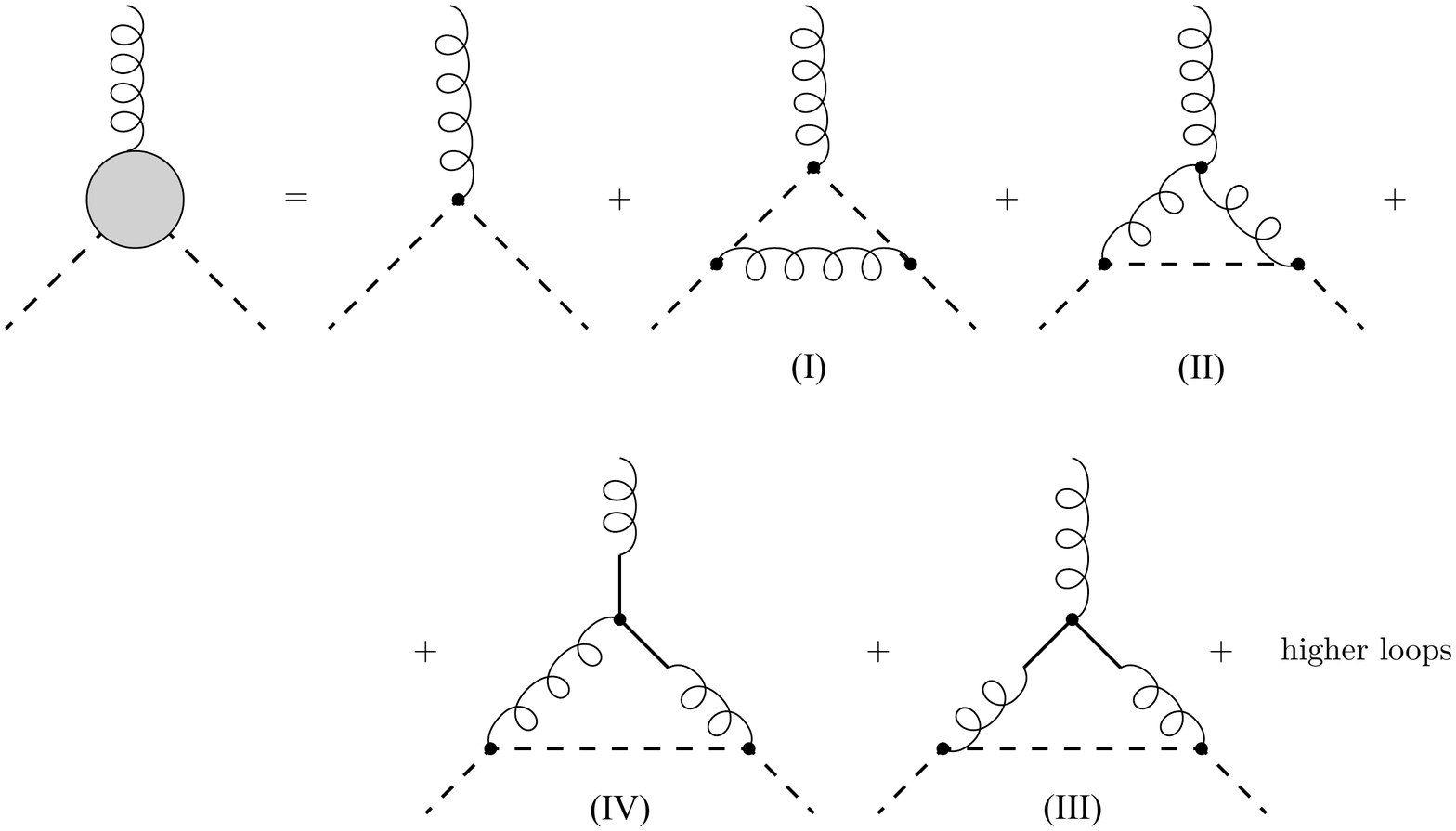}
\vspace{-.8cm}
\caption{Feynman diagram expansion up to one-loop order for the ghost-gluon vertex in the Refined Gribov-Zwanziger theory. Dashed lines represent ghosts and antighosts, while the curly lines stand for gluons. Full lines,  that only appear in mixed propagators, correspond to the auxiliary fields $\varphi,\bar\varphi$. The roman numbers identifying the one-loop diagrams will be used as reference in the appendices.}
\end{figure}

Before proceeding, let us remark that since the RGZ action 
contains bilinear couplings between fields, the theory contains 
mixed propagators, such as $\vev{A\varphi}$ and 
$\vev{A\bar\varphi}$. Therefore, the relation between 
connected and 1PI functions has to take such mixed propagators 
into account. This is made explicit in the Feynman diagrams of 
Fig. \ref{fig:feyndiags}. Such mixed propagators and vertices 
involving Zwanziger's auxiliary fields $\varphi$ and 
$\bar\varphi$ as well as their  fermionic counterparts 
$\omega$ and $\bar\omega$, arise as consequence of the local 
formulation of the Gribov horizon \footnote{Another 
equivalent possible formulation of the theory would include 
nonlocal, momentum-dependent vertices instead of the auxiliary 
fields. However, for the sake of using standard QFT 
techniques, we employ the  local version of the theory.}. We give 
further details for the interested reader in Appendix 
\ref{sec:appendix-mixed-propagators}.

The 1-loop connected function (\ref{eq:cbcA-def}) is then decomposed as
\begin{eqnarray}
 \vev{A_\mu^a(k)\,\bar c^b(p)\,c^c(q)} = G(p)G(q)D_{AA}(k)P_{\mu\nu}(k)\left\{\frac{\delta^3\Gamma}{\delta A_\nu^a(-k)\delta c^b(-p)\delta \bar c^c(-q)}
 +
 \frac{2g\gamma^2f^{ade}}{k^2+\mu^2}\frac{\delta^3\Gamma}{\delta c^b(-p)\delta \bar c^c(-q)\delta \varphi_\nu^{de}(-k)}\right\}_{q=-p-k}\nonumber\\
\end{eqnarray}
or, in a compact notation,
\begin{equation}\label{eq:Acc-local-shorthand-notation}
  \frac{\vev{A\,\bar c\, c}_c}{(\vev{\bar c\,c}_c)^2\vev{AA}_c} = \Gamma_{A\,\bar c\,c} + \frac{\vev{A\,\varphi}_c}{\vev{A\,A}_c}\Gamma_{\varphi\,\bar c\,c} 
  +\frac{\vev{A\,\bar\varphi}_c}{\vev{A\,A}_c}\Gamma_{\bar\varphi\,\bar c\,c}.
\end{equation}
There are clearly some differences between perturbative Yang-Mills and RGZ calculations 
of the vertex function. The first of them is the modification of the gluon propagator 
brought about by the restriction to the Gribov horizon, which can be 
understood as the appearance of a pair of generally complex conjugate poles.  A second difference is the presence of the tree-level 
$A\bar\varphi\varphi$ vertex, which couples the gluon to the auxiliary Zwanziger fields. 
This allows not only diagrams with auxiliary fields running in the internal loops, but also 
in the external legs, as long as the external propagator is a mixed one like, for example,  $\vev{A\varphi}$. This possibility is realized in (\ref{eq:Acc-local-shorthand-notation}), 
giving rise to the contributions $\Gamma_{\varphi\bar c c}$ 
and $\Gamma_{\bar\varphi\bar c c}$, not present in perturbative YM theory. Finally, note 
that these mixed contributions only appear from one-loop order onwards, as such vertices
are absent from the classical action (\ref{eq:RGZ-local-action}).

\subsection{The one-loop ghost-gluon vertex in the soft gluon limit}



As is well known, the full ghost-gluon vertex function has a nontrivial 
tensor structure (see, e.g., \cite{Ball:1980ax}). Given the many extra 
terms that the restriction to the Gribov horizon 
brings to the calculation, a full one-loop  evaluation of 
the vertex function demands in practice some automated algorithm, which will be deferred to future work. Here, we present an analytic calculation in the physically interesting {\it soft gluon 
limit}, {\it i.e.}  as the gluon momentum $k\rightarrow0$.

The diagrams contributing to the dressing of the ghost-gluon vertex up to one-loop order in the RGZ theory are displayed in Fig. \ref{fig:feyndiags}.
Diagrams in the first line are the ones which appear in perturbative YM one-loop calculations, while the second line displays two extra diagrams that appear in RGZ theory due to the presence of the auxiliary fields $\varphi,\bar{\varphi}$ that localize the Gribov horizon function. Since the gluon propagator is deeply altered in the infrared regime by the presence of the Gribov horizon, even the standard diagrams in the first line of Fig. \ref{fig:feyndiags}.
yield nonperturbative effects, dependent on the Gribov parameter and on 
the dimension-two condensates of the RGZ framework.

The details of the computation can be found in Appendix 
\ref{sec:appendix-1-loop-diagrams}. Here, we notice that, in the soft gluon limit, the contribution
containing a three-gluon vertex simplifies tremendously.
Besides, the 
$\Gamma_{\varphi\bar c c}$ and $\Gamma_{\bar\varphi\bar c c}$ 
kernels vanish  in this limit, yielding a vanishing diagram (IV) in 
Fig. \ref{fig:feyndiags}.
Finally, the $k\rightarrow0$ 
limit of the one-loop ghost-gluon kernel can be written as 
\begin{eqnarray}\label{eq:resultGamma}
 [\Gamma_{A\bar c c}^{(1)}(0,p,-p)]^{abc}_{\mu} 
&=&ig^3\frac{Nf^{abc}}{2}\bigg\{R_+ J_\mu(a_+;p) + R_- J_\mu(a_-;p)
+2R_+^2K_\mu(a_+,a_+;p) + 2R_-^2K_\mu(a_-,a_-;p) + \nonumber\\
&&\left.+4R_+R_- K_\mu(a_+,a_-;p) +
\frac{N}{2}\left(\frac{g\gamma^2}{a_+^2-a_-^2}\right)^2
\left[K_\mu(a_+,a_+;p)+K_\mu(a_-,a_-;p)-\right.\right.\nonumber\\
&&-2K_\mu(a_+,a_-;p)\big]\bigg\}\,,
\end{eqnarray}
where the master integrals
\begin{equation}
 J_\mu(m_1;p):=\int_\ell\,
\frac{1}{\ell^2}\frac{1}{\ell^2+m_1^2}\frac{p^2\ell^2 - (p\cdot\ell)^2}{[(\ell-p)^2]^2}\,(\ell-p)_\mu,
\end{equation}
related to diagram (I),
and
\begin{eqnarray}
 K_\mu(m_1,m_2;p)&:=&\int_\ell \frac{1}{(\ell+p)^2}\frac{1}{\ell^2+m_1^2}\frac{1}{\ell^2+m_2^2}
\,\left[\frac{\ell^2p\cdot p - (p\cdot\ell)^2}{\ell^2}\right]\ell_\mu\,,
\end{eqnarray}
which appears in diagrams (II) and (III),
have been explicitly calculated in the Appendix \ref{sec:appendix-1-loop-diagrams}. The incoming 
antighost momentum is given by $p$.  Therefore, the ghost momentum is $-p$, since $k=0$. The massive parameters
$-a_\pm^2$ are the,  generally complex, poles of the RGZ gluon propagator (\ref{eq:RGZgluonpropagator}) and 
$R_\pm$ are their corresponding residues. 
It is interesting to point out that the last terms in eq. \eqref{eq:resultGamma} come from new diagram (III) in Fig. \ref{fig:feyndiags} 
which is absent in standard YM theories, being completely nonperturbative and proportional to $\gamma^4$. Note that the integrals are also valid for complex arguments.

Before proceeding to the numerical analysis of the next section, it is important to remark that the one-loop 
vertex function explicitly respects the so-called Taylor kinematics, {\it i.e.}, 
\begin{eqnarray}\label{eq:Taylor-kin-antighost}
 (\Gamma_{A\,\bar c\,c})^{abc}_\mu(p,0,-p) = 0\,,
\end{eqnarray}
and the so-called non-renormalization theorem of the ghost-gluon vertex, namely 
\begin{eqnarray}\label{eq:Taylor-kin-ghost}
 (\Gamma_{A\,\bar c\,c})^{abc}_\mu(-p,p,0) = -i gf^{abc}p_\mu,
\end{eqnarray}
which are the same in the RGZ framework as in perturbative 
Yang-Mills theory \cite{Taylor:1971ff}. These are direct
consequences of the Ward identities of the action 
(\ref{eq:RGZ-local-action}), as shown in Appendix \ref{sec:appendix-ward}.

\section{Results and discussion} 
\label{sec:discussion}

The final result for the one-loop correction of the gluon-ghost vertex in the soft-gluon limit in the RGZ theory is given in Eq.\eqref{eq:resultGamma} as a function of the poles $a_{\pm}$ and residues $R_{\pm}$ of the tree-level gluon propagator, as well as of the Gribov parameter $\gamma$. The gluon propagator in the RGZ theory is modified with respect to standard YM, even at tree level, by the presence of the Gribov parameter and of dimension-two condensates of the gluon and auxiliary fields. 
The gluon dressing function in $d=4$ takes the form:
\begin{eqnarray}
 D(p^2) = \frac{p^2+M^2}{p^4+(M^2+m^2)p^2 + M^2m^2+2g^2N\gamma^4}\equiv\frac{p^2+a}{p^4+bp^2+c}\,,
\end{eqnarray}
where $M$ and $m$ are mass parameters related to dimension-two condensates, summing up to the total of three parameters ($M,m,\gamma$) in the RGZ theory, besides the gauge coupling $g$ and possible renormalization scale.

The self-consistency of the RGZ theory allows one in principle to compute all of these three parameters, using the Gribov gap equation, renormalization group invariance and a minimization of the RGZ effective potential. The calculation of these parameters can however only be done up to a certain order of approximation and involves lengthy analyses that are not the aim of the current work. For more details on how to proceed in these lines, the reader is referred to the review in Ref. \cite{Vandersickel:2011zc} and references therein. 

We shall proceed to make quantitative predictions and comparisons with other results in the literature by fixing the three RGZ parameters through a fit of the lattice YM data for the gluon propagator with the tree-level form obtained in the RGZ theory. As discussed in the introduction, this type of fit works remarkably well for low and intermediate momenta \footnote{For large momenta, perturbative logarithmic corrections must be added to this tree-level expression in order to have good agreement with lattice data.}. This success is reassuring in the sense that the RGZ theory might indeed be capturing a significant fraction of the nonperturbative phenomena of infrared YM theory. The current one-loop analysis of the ghost-gluon vertex goes in the direction of further verifying how much of the nonperturbative YM correlations may be described by the RGZ theory to a reasonably low order in perturbation theory.

\subsection{Parameter fixing}

In the next subsections, results for the ghost-gluon vertex for the SU(2) and SU(3) cases will be presented. Each non-Abelian gauge group gives rise to a different parameter set from the corresponding fits of the lattice gluon propagator. For SU(2) we use the fit displayed in Fig.2 of Ref.\cite{Cucchieri:2011ig}, corresponding to the largest volume ($V=128^4$) data set and improved momenta. The SU(3) parameters are given in Ref. \cite{Oliveira:2012eh} (cf. their Fig.7; $\beta=6.2$) and include an infinite-volume extrapolation. The parameter sets are summarized in Table \ref{table:parameters}\footnote{Note that the fits presented in the cited lattice references include an overall renormalization factor as an extra fit parameter. In our RGZ model, this factor is fixed at $Z=1$ at this order of perturbation theory. Nevertheless, the fixing of this overall constant can be seen as a renormalization condition for the propagator.}.

\begin{table}[h!]
  \begin{center}
    \begin{tabular}{c||c|c|c}
     Gauge group & \hspace{.2cm} $M^2$ (${\rm GeV}^2$) \hspace{.2cm} & \hspace{.2cm} $m^2$ (${\rm GeV}^2$) \hspace{.2cm} & \hspace{.2cm} $2g^2N\gamma^4$ (${\rm GeV}^4$) \hspace{.2cm} \rule{0pt}{3ex} \\
\hline\hline
\rule{0pt}{3ex} 
 SU(2) - Ref.\cite{Cucchieri:2011ig}  & 2.508(78)    & $[0.768(17)]^2$ &   $[0.720(9)]^2$\\
 \rule{0pt}{2ex} 
 SU(3) - Ref.\cite{Oliveira:2012eh}&   4.473(21)  & 0.704(29)   & 0.3959(54)
\end{tabular}  
\end{center}
    \caption{RGZ parameters fitted from lattice results for SU(2) \cite{Cucchieri:2011ig} and SU(3) \cite{Oliveira:2012eh}, with error estimates in parenthesis.}
    \label{table:parameters}
\end{table}

Since all integrals are finite (see Ap.~\ref{sec:appendix-1-loop-diagrams}) and,  therefore, there is no explicit renormalization scale dependence in the results, the only missing free parameter is the coupling constant $g$.
Results will be shown for different values of $g$, within the perturbative range, and also for a running coupling corresponding to the standard one-loop YM beta function in the Modified Minimal Subtraction scheme, i.e. \cite{Peskin:1995ev}
\begin{eqnarray}\label{Eq:Running}
 g^2(p) = \frac{g^2(\mu)}{1+\frac{11N}{3}\,\frac{g^2(\mu)}{8\pi^2}\log(\frac{p}{\mu})}\,.
\end{eqnarray}
%

\subsection{SU(2) case}

Let us now present the results for the SU(2) ghost-gluon vertex form factor in the soft gluon limit, in which the gluon momentum is taken to zero. The full vertex tensor in this limit becomes:
\begin{eqnarray}
 [\Gamma_{A\bar c c}(0,p,-p)]^{abc}_{\mu} 
&=&-ig f^{abc} p_{\mu}\Gamma_{A\bar c c}(p)
\,,
\end{eqnarray}
where the form factor $\Gamma_{A\bar c c}(p)$ is the scalar function of the momentum which shall be analyzed in what follows.

In general, the antighost-momentum dependence of the effect of interactions on the vertex 
is that of rising from the tree-level value at $p=0$ reaching a peak around $p\sim 1$ GeV and slowly falling again at very large momentum, eventually saturating at the perturbative result for fixed coupling 
\begin{eqnarray}
\Gamma_{A\bar c c}(p)= 1+ g^2N \frac{3}{64\pi^2}
\,,
\end{eqnarray}
which corresponds exactly to the infinite momentum limit of Eq. \eqref{eq:resultGamma}.
Therefore, interactions are consistently suppressed in the deep ultraviolet as expected from asymptotic freedom, which remains untouched in the RGZ theory as already proven via algebraic renormalization analyses \cite{Dudal:2008sp}. Moreover, the nonmonotonic momentum dependence observed as contrasted to the flatness of the one-loop perturbative result is a sign of the nonperturbative nature of the current analysis. In fact, these properties are consistently found in all nonperturbative methods that we compare to here -- from Curci-Ferrari model and Dyson-Schwinger equations to lattice simulations \footnote{Due to large error bars in lattice data, it is not possible to exclude a crossing below the tree level value for low momenta.} -- as well as other approaches (cf. e.g. \cite{Cyrol:2016tym}).

\begin{figure}[h!]
\includegraphics[height=6cm]{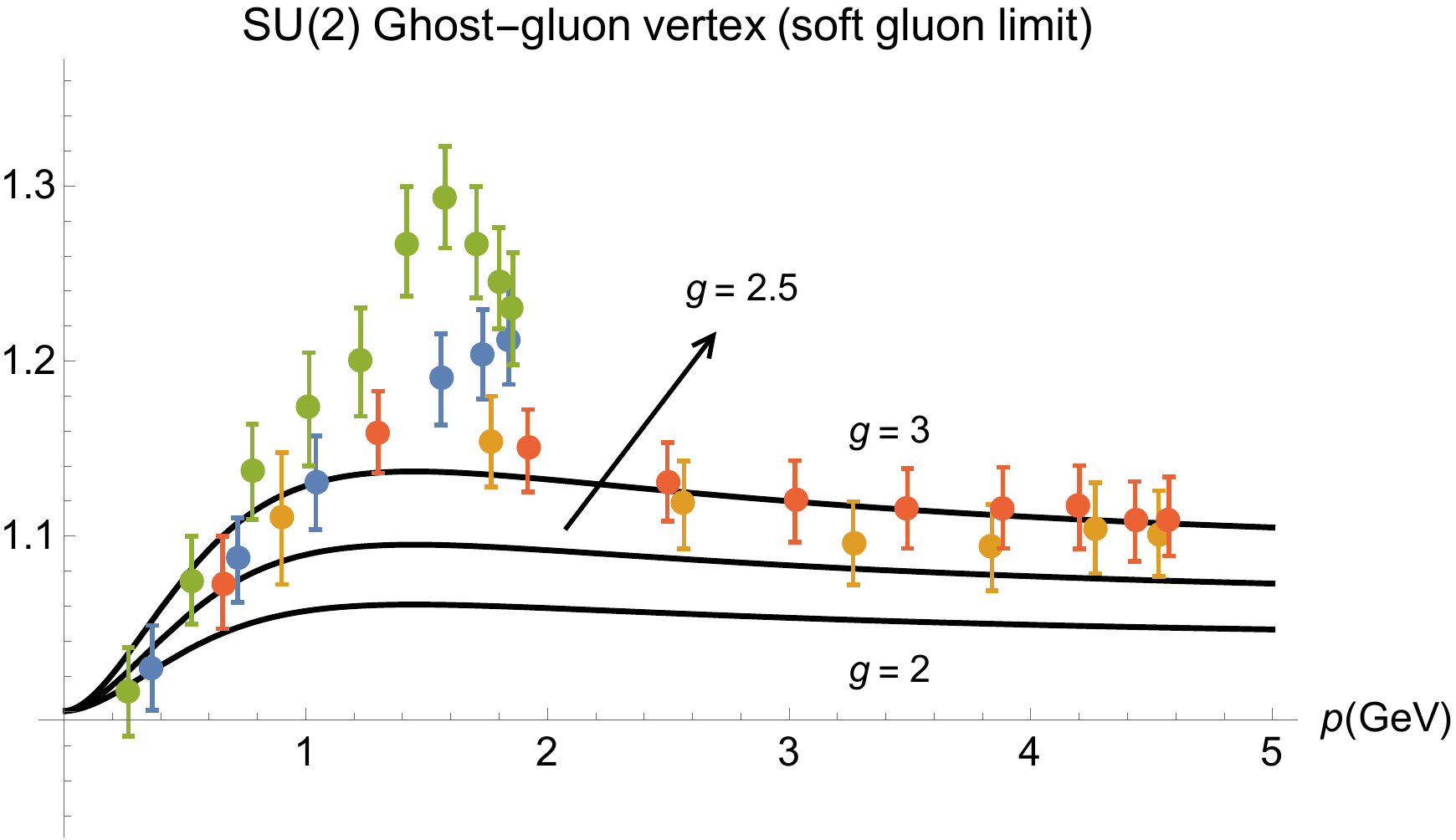}
\caption{The form factor $\Gamma_{A\bar c c}(p)$ of the SU(2) ghost-gluon vertex in the soft-gluon limit as a function of the antighost momentum for $d=4$ is compared to lattice simulations. The solid lines 
represent our results for different values of the coupling: $g=2,\,  2.5, \, 3$, from the bottom to the top curve, respectively. Data points correspond to lattice results from Ref. \cite{Cucchieri:2008qm}.
}\label{fig:SU(2)compLattice}
\end{figure}

Figure \ref{fig:SU(2)compLattice} displays our results,  i.e. Eq.(\ref{eq:resultGamma}) with the parameters from the SU(2) line in Table \ref{table:parameters}, as compared to lattice data from Ref. \cite{Cucchieri:2008qm}. We plot the form factor of the ghost-gluon vertex here for three different values of the coupling $g$. For momenta below $\sim 1$ GeV and above $\sim 2$ GeV the RGZ results with a coupling of $g=2.5 - 3$ GeV provide a reasonable description of the available lattice data. It is interesting to point out that these values of coupling correspond to $\alpha=\frac{g^2N}{12\pi}<0.5$, being in principle within a regime of applicability of the perturbative approximation. In the region of intermediate momenta ($p=1-2$ GeV), there is an apparent disagreement between different lattice data sets and improved simulations on larger lattices with more statistics are probably needed to resolve this issue.

\begin{figure}
\includegraphics[height=6cm]{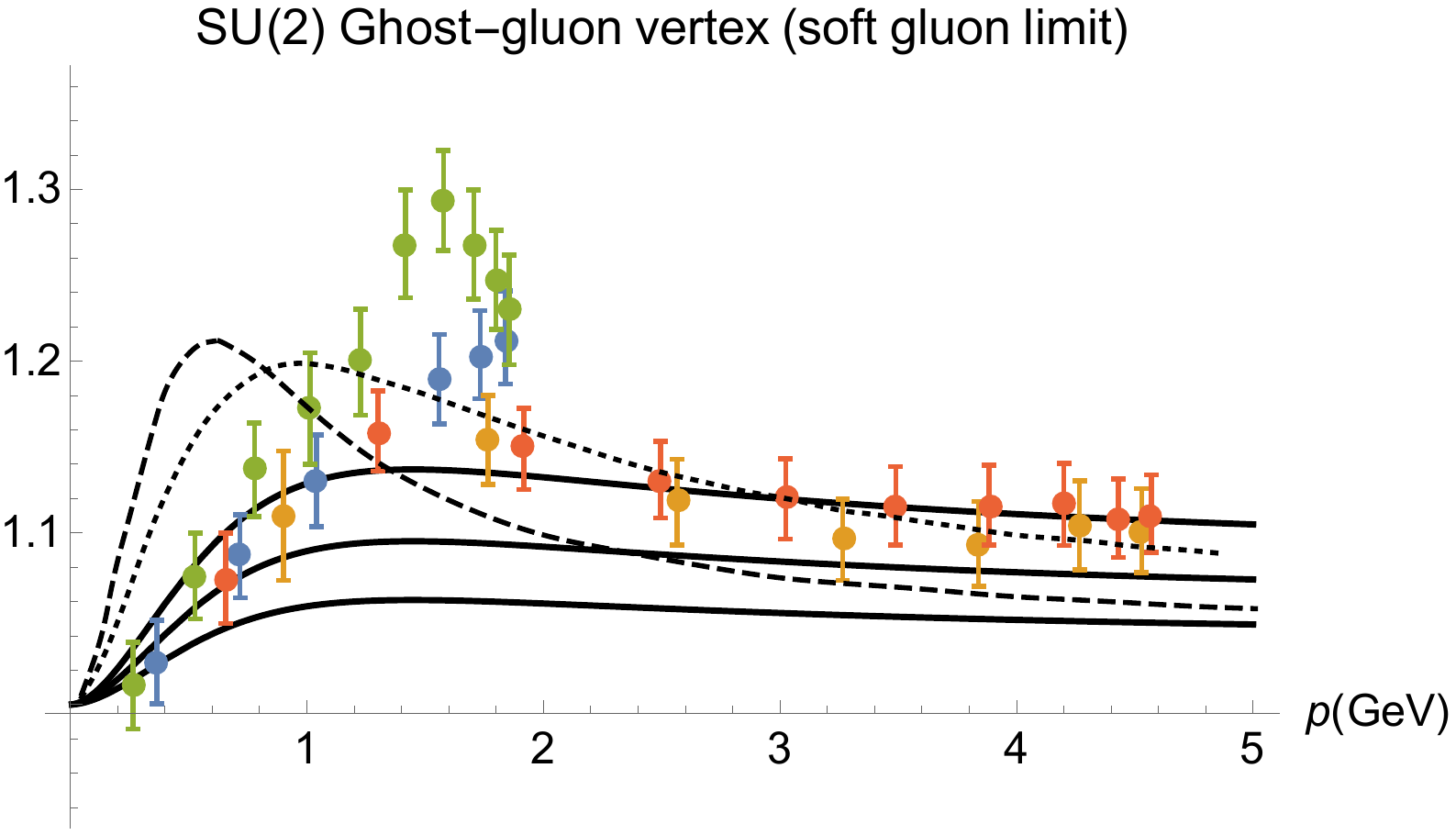}
\caption{The scalar function $\Gamma_{A\bar c c}(p)$ of the SU(2) ghost-gluon vertex in the soft-gluon limit as a function of momentum for $d=4$. The solid lines 
represent our results for different values of the coupling: $g=2,\,  2.5, \, 3$, from the bottom to the top curve, respectively. Perturbative calculations with RG-improvement in the Curci-Ferrari model
\cite{Pelaez:2013cpa} are displayed as the dashed and the dotted lines, which correspond to different renormalization schemes. Lattice data points from Ref. \cite{Cucchieri:2008qm}.
}\label{fig:SU(2)compPelaez}
\end{figure}

One can further compare the RGZ vertex results with the outcome of different nonperturbative methods. In Figure \ref{fig:SU(2)compPelaez} findings for the SU(2) ghost-gluon vertex in the soft-gluon limit in the renormalization-group (RG) improved Curci-Ferrari model at one-loop order \cite{Pelaez:2013cpa} are added in the comparison. The qualitative behavior is the same, but the peak  
is more pronounced and shifted towards lower momenta; the intensity of those effects being dependent on the renormalization scheme adopted. 
The stronger fall of the correlator for large momenta may be seen as the direct effect of the running coupling due to asymptotic freedom, which is absent from our fixed-$g$ curves. In the SU(3) case below, we shall discuss a na\"\i ve inclusion of RG corrections in our RGZ one-loop vertex results that will show exactly this property. 

It should also be  noted that the RG schemes adopted in \cite{Pelaez:2013cpa} are nonstandard, with the absence of a Landau pole being an important feature at this one-loop implementation. Since the Curci-Ferrari model includes a mass term for the gluon, the RG flow will in general involve coupled equations for the coupling, the mass and the field renormalizations. Choosing a convenient scheme, with a mass-dependent beta function, the authors of \cite{Pelaez:2013cpa} (cf. also \cite{Tissier:2011ey}) provide a fully smooth behavior for the RG-improved correlation functions down to zero momentum.

\subsection{SU(3) case}

\begin{figure}[h!]
\includegraphics[height=6cm]{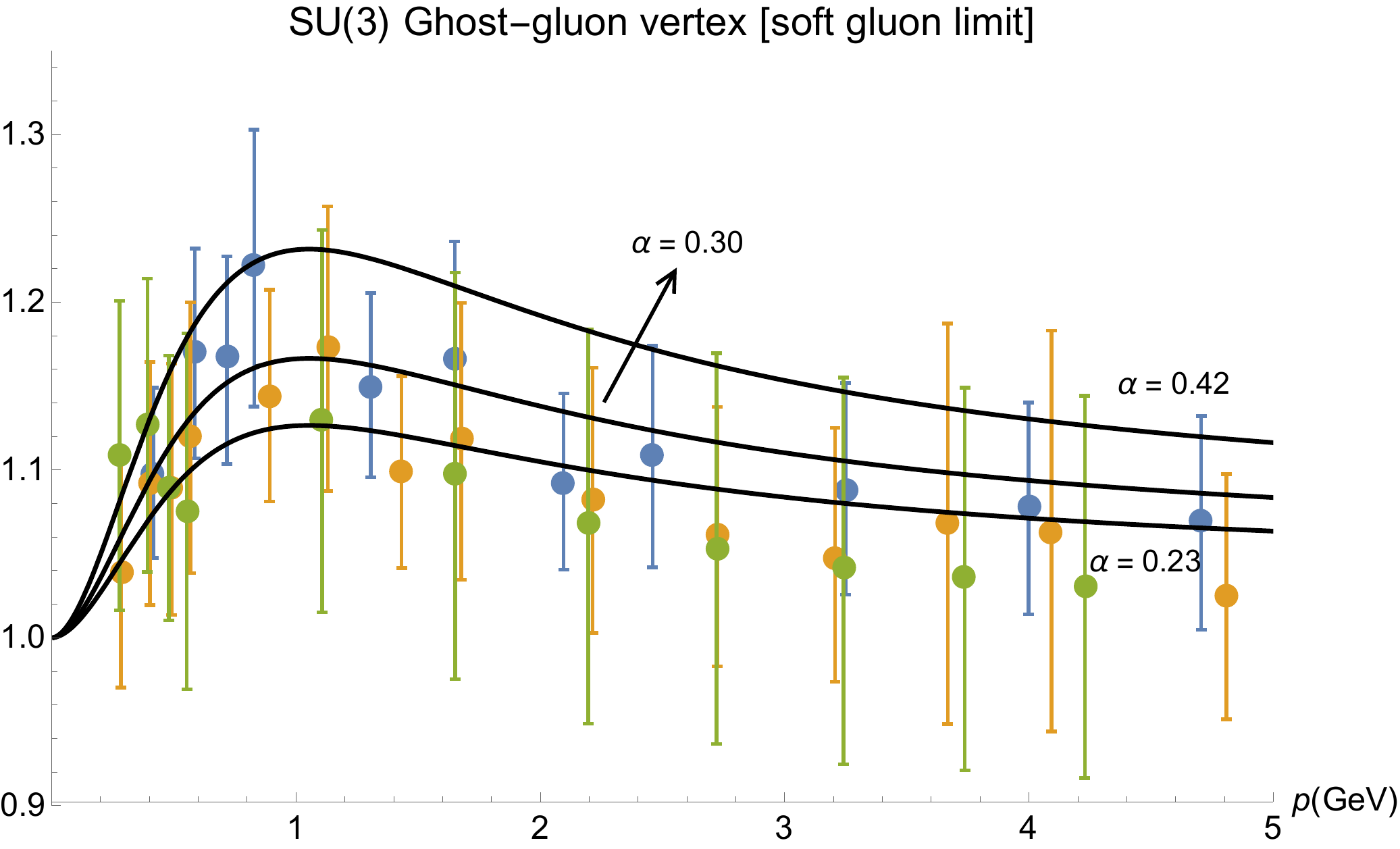}
\caption{The form factor $\Gamma_{A\bar c c}(p)$ of the SU(3) ghost-gluon vertex in the soft-gluon limit as a function of the antighost momentum for $d=4$ is compared to lattice simulations. The solid lines 
represent our results for different values of the strong coupling: $\alpha=\frac{g^2}{4\pi}=0.23,\, 0.3, \, 0.42$, from the bottom to the top curve, respectively. Data points correspond to lattice results from Ref. \cite{Ilgenfritz:2006he}.
}\label{fig:SU(3)compLattice}
\end{figure}

In the SU(3) case, we adopt the parameters fitted from lattice propagators in the last line of Table \ref{table:parameters}. The form factor of the ghost-gluon vertex is again plotted as a function of the antighost momentum $p$ in Figure \ref{fig:SU(3)compLattice}, for fixed coupling $\alpha=\frac{g^2}{4\pi}=0.23,\, 0.3,\, 0.42$, and compared to lattice data \cite{Ilgenfritz:2006he}.

Qualitatively, the same peak structure around $p\sim 1$ GeV is also observed in the SU(3) case. Moreover, Figure \ref{fig:SU(3)compLattice} shows that the RGZ results are once again compatible with the available SU(3) lattice data for a large range of (fixed) values of the coupling, $\alpha=0.23-0.42$, falling nicely within a perturbative domain, so that our one-loop approximation in RGZ theory is in principle consistent.

\begin{figure}
\includegraphics[height=6cm]{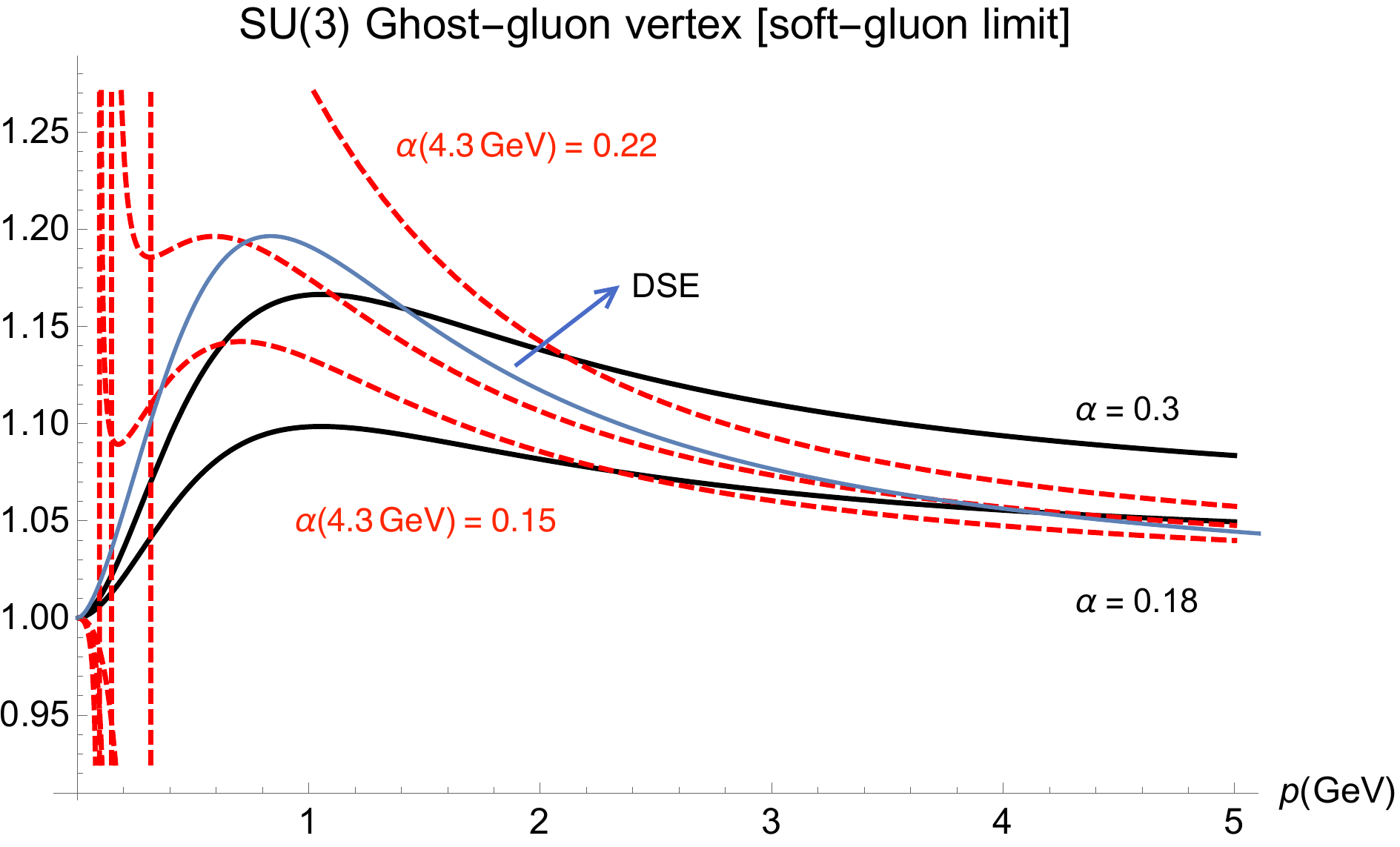}
\caption{The scalar function $\Gamma_{A\bar c c}(p)$ of the SU(2) ghost-gluon vertex in the soft-gluon limit as a function of momentum for $d=4$.
Solid thick (black) lines represent our results for different values of the strong coupling: $\alpha=\frac{g^2}{4\pi}=0.18$ (bottom) and $\alpha=0.3$ (top), while the dashed (red) lines include
a one-loop perturbative running coupling with different renormalization conditions: $\alpha(\mu=4.3 {\rm GeV})=0.15,\,  0.18, \, 0.22$, from the bottom to the top curve, respectively. 
A DSE result from Ref. \cite{Aguilar:2013xqa} is represented as the thin (blue) line. }\label{fig:SU(3)compDSErunning}
\end{figure}

For SU(3) there are dynamical solutions of the Dyson-Schwinger equations (DSE) for the ghost-gluon vertex in the soft gluon limit available within different truncation schemes in the literature (cf. e.g. \cite{Aguilar:2013xqa,Huber:2012kd}).
Overall, the qualitative behavior is the same one observed here and a quantitative comparison with one of these DSE results is shown in Figure \ref{fig:SU(3)compDSErunning}. For antighost momenta above $\sim 1$ GeV one has a steeper fall of the form factor in the DSE case as compared to our fixed-coupling results. If we consider asymptotic freedom, it is reasonable to assume that the coupling will indeed decrease for large momenta, giving thus rise to the stronger suppression of the ghost-gluon vertex. The dashed (red) lines in Figure \ref{fig:SU(3)compDSErunning} show that a na\"\i ve implementation of the perturbative running coupling (cf. \eqref{Eq:Running}) generates a steeper fall of the vertex form factor for large momenta, being very close to the one observed in the DSE solution \cite{Aguilar:2013xqa}. One may conclude thus that the quantitative difference for large momenta between the RGZ results and dynamical DSE solutions for the ghost-gluon vertex is naturally accounted for by the effect of the running coupling.
For small momenta, non-analytical behavior is observed, being related to the presence of a Landau pole in the one-loop perturbative running adopted.

This na\"\i ve implementation of the running coupling is however not the full RG flow in RGZ theories. Due to the existence of extra operators -- related to the Gribov horizon and the dimension-two condensates --, the RG flow in RGZ theory corresponds to a set of coupled equations for the running coupling as well as the running massive parameters $m,M,\gamma$. Even though we shall not attempt here to derive the full flow, it is in principle feasible and could be done in an infrared safe scheme, similar to the one developed in \cite{Tissier:2011ey} and free of Landau poles.

\section{Final remarks} 
\label{sec:finalremarks}
 
Nonperturbative descriptions of the infrared regime of  YM theories are crucial to understand the physics of confinement 
in QCD. Among several available approaches, the refined Gribov-Zwanziger framework partially solves the problem of Gribov ambiguities in the gauge path integral and provides a description that reproduces perturbative YM in the ultraviolet regime, while displaying nontrivial infrared physics via the Gribov horizon background. 

The RGZ theory has been successfully tested against the nonperturbative benchmark of lattice data for two-point correlation functions and has provided reasonable estimates for observables like e.g. the glueball mass spectra. Up to now most of these applications have considered the leading-order perturbative approximation and have not explored higher-order correlation functions. In this paper, we compute the one-loop ghost-gluon vertex in the RGZ theory. The trivial result in the Taylor kinematics (vanishing antighost momentum), as well as the tree-level vertex when the ghost momentum goes to zero, have both been reproduced and shown to be direct consequences of Ward identities of the RGZ action. Moreover, we obtain an analytical result for the one-loop ghost-gluon vertex in the limit of vanishing gluon momentum. This calculation is an important test of the capability of the RGZ approach to provide a consistent nonperturbative description of YM theories that goes beyond two-point correlation functions.

Our findings for both SU(2) and SU(3) cases qualitatively agree with other nonperturbative approaches, such as dynamical DSE solutions in different truncation schemes \cite{Aguilar:2013xqa,Huber:2012kd} and the RG-improved Curci-Ferrari model \cite{Pelaez:2013cpa}. Quantitative differences at large momenta have been shown to be accounted for by the running of the strong coupling and asymptotic freedom. The RGZ one-loop ghost-gluon vertex is also quantitatively compatible with the available lattice data for SU(2) \cite{Cucchieri:2008qm} and SU(3) \cite{Ilgenfritz:2006he} gauge groups in the soft-gluon limit, even for fixed coupling.
It is important to note that the momentum behavior of the form factor of the ghost-gluon vertex is not fitted. All massive parameters of the RGZ theory are fixed by lattice data for the gluon propagator; the only free parameter in the vertex correction being an overall factor $g^2$, where $g$ is the gauge coupling. In general, the set of results for the ghost-gluon vertex provides further indication that indeed the RGZ action may be seen as a nonperturbative infrared description of YM dynamics.

Even though the results are already encouraging, there are many improvements that could be done in the current analysis. A fully dynamical calculation of the RGZ dimension-two condensates (i.e. the massive parameters $m,M$) would allow one to have a self-consistent prediction of the theory. One step further with respect to the calculation presented here would be to implement the full RG flow given by the set of coupled equations of the massive parameters, possibly in an infrared safe scheme. Another interesting analysis made possible by 
the action (\ref{eq:RGZ-local-action}) is the study of 
the gauge-parameter dependence -- within linear covariant gauges -- of the gluon-ghost vertex
we have just investigated in the Landau gauge.


\section*{Acknowledgements}

The authors acknowledge useful discussions with M. Tissier and N. Wschebor. The authors also would like to thank A. C. Aguilar and A. Maas for discussions and for kindly providing the data for the comparison plots in Section \ref{sec:discussion}.
This work has been partially supported by CNPq, CAPES, and 
FAPERJ. It is a part of the project INCT-FNA Proc. 
464898/2014-5. A.D.P acknowledges funding by the DFG, Grant Ei/1037-1.

\appendix 

\section{Relevant Feynman rules (Landau gauge)}\label{sec:appendix-feynman-rules}

Given its many fields and interactions, the RGZ theory has a large number of propagators and vertices. 
However, for the calculation of the ghost-gluon vertex at one-loop level (and in the Landau gauge), only a few 
of them are required. The Feynman rules corresponding to these propagators and vertices are shown below.

\subsection{Tree-level propagators}

In order to calculate the ghost-antighost-gluon 3-point function in the 
Refined Gribov-Zwanziger theory, only a subset of the propagators of 
the theory are needed. These are 
\begin{eqnarray}
\langle A^a_{\mu}(p)A^b_{\nu}(-p)\rangle &=& \delta^{ab}\left[\frac{p^2+M^2}{p^4+(m^2+M^2)p^2+m^2M^2+2Ng^2\gamma^4}
\mathcal{P}_{\mu\nu}(p)\right]\equiv \delta^{ab}\mathcal{P}_{\mu\nu}(p)D_{AA}(p)\label{eq:RGZgluonpropagator}\\\nonumber
\\
\langle A_{\mu}^a(p)\varphi_{\nu}^{bc}(-p)\rangle
&=&\frac{g\gamma^2f^{abc}}{p^4+p^2(m^2+M^2)+m^2M^2+2Ng^2\gamma^4}\mathcal{P}_{\mu\nu}(p) 
= g\gamma^2f^{abc}\mathcal{P}_{\mu\nu}(p)\frac{D_{AA}(p)}{p^2+M^2}\\
\langle A_{\mu}^a(p)\bar{\varphi}_{\nu}^{bc}(-p)\rangle
&=&-\frac{g\gamma^2f^{abc}}{p^4+p^2(m^2+M^2)+m^2M^2+2Ng^2\gamma^4}\mathcal{P}_{\mu\nu}(p)=-\langle A_{\mu}^a(p)\varphi_{\nu}^{bc}(-p)\rangle\\
\langle\bar{c}^a(p)c^b(-p)\rangle &=&\frac{1}{p^2}\delta^{ab}\equiv \delta^{ab}D_{\bar cc}(p)
\;
\end{eqnarray}
where
\begin{eqnarray}
 \mathcal{P}_{\mu\nu}(p) = \delta_{\mu\nu} - \frac{p_\mu p_\nu}{p^2}
\end{eqnarray}
is the transverse projector, such that $p_\mu\mathcal{P}_{\mu\nu}(p)=p_\nu\mathcal{P}_{\mu\nu}(p)=0$. 

\subsection{Tree-level vertices}

The only vertices needed for the computation of the ghost-antighost-gluon 
at one-loop in the RGZ theory are
\begin{eqnarray}
 ^{tree}[\Gamma_{AAA}(k,p,q)]^{abc}_{\mu\nu\rho}&=&-\left.\frac{\delta^3S_{tree}}{\delta A_\mu^a(k)\delta A_\nu^b(p)\delta A_\rho^c(q)}\right|_{\Phi=0} 
 = igf^{abc}\left[(k_\nu-q_\nu)\delta_{\rho\mu} + (p_\rho-k_\rho)\delta_{\mu\nu} + (q_\mu-p_\mu)\delta_{\nu\rho} \right] \nonumber\\
^{tree}[\Gamma_{A\bar c c}(k,p,q)]^{abc}_{\mu}&=& -\left.\frac{\delta^3S_{tree}}{\delta A_\mu^a(k)\delta \bar c^{b}(p) \delta c^{c}(q) }\right|_{\Phi=0} = -igf^{abc}p_\mu\nonumber\\
 ^{tree}[\Gamma_{A\bar \varphi \varphi}(k,p,q)]^{abcde}_{\mu\nu\rho}&=& -\left.\frac{\delta^3S_{tree}}{\delta A_\rho^a(k)\delta \bar\varphi_\mu^{bc}(p) \delta\varphi_\nu^{de}(q) }\right|_{\Phi=0} = -igf^{abd}\delta^{ce}\delta_{\nu\rho}p_\mu \nonumber\\
\end{eqnarray}

Note that, for higher orders in the perturbative expansion, or for a general 
linear covariant gauge, or for other correlation functions, extra correlators 
and vertices will be needed.

\section{On the relation between connected and 1PI correlation functions in the presence of mixed propagators}\label{sec:appendix-mixed-propagators}

Let us denote the generating 
functional of connected correlation functions as $W[\vec J]$, where $J_i$ are external sources associated 
with the different elementary fields, and let $\Gamma[\vec\phi]$ be the quantum action, that is, the 
generating functional of 1PI correlation functions. Using this notation, we start from the well-known 
relation
 \begin{eqnarray}\label{eq:2point-connected-1PI-J}
\left.\frac{\delta^2\Gamma[\vec\phi]}{\delta\phi_j\delta\phi_\ell}\right|_{\vec \phi = \vec{\Phi}[\vec J]}\frac{\delta^2W[\vec J]}{\delta J_\ell\delta J_k} 
 = -\delta_{jk} 
\end{eqnarray}
Taking a further derivative with respect to the source $J_i$, one finds
\begin{eqnarray}\label{eq:connected-3-point}
 \frac{\delta^3W[\vec J]}{\delta J_i\delta J_p\delta J_k}&=& 
 -\frac{\delta^2W[\vec J]}{\delta J_p\delta J_j}\left(\left.\frac{\delta^3 
 \Gamma[\vec \phi]}{\delta \phi_j\delta \phi_\ell\delta\phi_m}\right|_{\vec \phi = \vec{\Phi}[\vec J]}\right)
 \frac{\delta^2W[\vec J]}{\delta J_i\delta J_m}\frac{\delta^2W[\vec J]}{\delta J_\ell\delta J_k}.
\end{eqnarray}

For the present calculation of the ghost-gluon vertex, we are specifically interested in the choice
\begin{eqnarray}
  i &=& A_\mu^e(k)\nonumber\\
  p &=& \bar c^a(p)\nonumber\\
  k &=& c^b(q).
\end{eqnarray}

Since there are no mixed propagators involving the Faddeev-Popov ghosts $c$ and $\bar c$, the only nonvanishing 
contributions are such that $j=c$ and $\ell=\bar c$. Therefore, running the remaining sum for 
$m=A,\varphi,\bar\varphi$, we have 
 \begin{eqnarray}
  \frac{\delta^3W[\vec J]}{\delta J_A\delta J_{\bar c}\delta J_c}&=& 
 -\frac{\delta^2W[\vec J]}{\delta J_c\delta J_{\bar c}}
 \left\{\left(\left.\frac{\delta^3 \Gamma[\vec \phi]}{\delta c\,\delta {\bar c}\,\delta A}\right|_{\vec \varphi = \vec{\Phi}[\vec J]}\right)
 \frac{\delta^2W[\vec J]}{\delta J_A\delta J_A} + \right.\nonumber\\
 &&+\left.\left(\left.\frac{\delta^3 \Gamma[\vec \phi]}{\delta c\,\delta \,{\bar c}\,\delta\varphi}\right|_{\vec \varphi = \vec{\Phi}[\vec J]}\right)
 \frac{\delta^2W[\vec J]}{\delta J_A\delta J_\varphi} + \right.
  \left.\left(\left.\frac{\delta^3 \Gamma[\vec \phi]}{\delta c\,\delta \,{\bar c}\,\delta\bar\varphi}\right|_{\vec \varphi = \vec{\Phi}[\vec J]}\right)
 \frac{\delta^2W[\vec J]}{\delta J_A\delta J_{\bar\varphi}}\right\}
 \frac{\delta^2W[\vec J]}{\delta J_{\bar c}\delta J_c},
 \end{eqnarray}
which can be written as
\begin{eqnarray}
 \vev{\bar c^a(p)\,c^b(q)\,A_\mu^c(k)} = D_{\bar cc}(p)D_{\bar cc}(q) D_{AA}(k)P_{\mu\nu}(k)\left\{\frac{\delta^3\Gamma}{\delta c^a(-p)\delta \bar c^b(-q)\delta A_\nu^c(-k)}
 +
 \frac{2g\gamma^2f^{cde}}{k^2+\mu^2}\frac{\delta^3\Gamma}{\delta c^a(-p)\delta \bar c^b(-q)\delta \varphi_\nu^{de}(-k)}\right\}\nonumber\\
\end{eqnarray}
or, in a shorthand notation,
 \begin{equation}\label{eq:Acc-local}
  \frac{\vev{A\,\bar c\, c}_c}{(\vev{\bar c\,c}_c)^2\vev{AA}_c} = \Gamma_{A\,\bar c\,c} + \frac{\vev{A\,\varphi}_c}{\vev{A\,A}_c}\Gamma_{\bar c\,c\,\varphi} 
  +\frac{\vev{A\,\bar\varphi}_c}{\vev{A\,A}_c}\Gamma_{\bar c\,c\,\bar\varphi}.
 \end{equation}

Therefore, besides the contribution $\Gamma_{A\bar cc}$, that would be present at pure Yang-Mills, there are also contributions from 1PI functions involving the auxiliary fields $\varphi$ and $\bar\varphi$, as well as the respective mixed propagators.

\section{Symmetries of the action and Ward identities}\label{sec:appendix-ward}

In this appendix, we wish to establish important simplifying 
relations between correlation functions in the RGZ framework. 
With this in mind, let us consider its symmetries and the 
corresponding Ward identities. First, let us notice that the 
RGZ action (\ref{eq:RGZ-local-action}) is left invariant by 
the following nilpotent BRST transformations 
\cite{Capri:2016aqq}
\begin{align}
sA^{a}_{\mu}&=-D^{ab}_{\mu}c^b\,,     &&sc^a=\frac{g}{2}f^{abc}c^bc^c\,, \nonumber\\
s\bar{c}^a&=b^{a}\,,     &&sb^{a}=0\,, \nonumber\\
s\varphi^{ab}_{\mu}&=0\,,   &&s\omega^{ab}_{\mu}=0\,, \nonumber\\
s\bar{\omega}^{ab}_{\mu}&=0\,,         &&s\bar{\varphi}^{ab}_{\mu}=0\,,\nonumber \\
s h^{ij}& = -ig c^a (T^a)^{ik} h^{kj}  \;, && sA^{h,a}_\mu =0\,,   \nonumber \\
s\tau^a& =0\,, &&  
\label{npbrst16a}
\end{align}
from which the BRST transformation of the field $\xi^a$, 
Eq.(\ref{eq:def-h}), can be evaluated iteratively, leading to
\begin{eqnarray}
s \xi^a&=&  - c^a + \frac{g}{2} f^{abc}c^b \xi^c - \frac{g^2}{12} f^{amr} f^{mpq} c^p \xi^q \xi^r + O(g^3)    \;\nonumber\\
&\equiv&g^{ab}(\xi)c^b.
\label{eq:sxi}
\end{eqnarray}
Such relations allow one to demonstrate the important properties
\begin{eqnarray}
s^2&=&0\;,\nonumber\\
s  S^{\mathrm{loc}}_{\mathrm{RGZ}} &=& 0   \;. \label{brstex}
\end{eqnarray}

The BRST invariance of the action is particularly important since it leads to the 
independence of important quantities (such as the poles of the $\vev{AA}$ 
correlator, or the Gribov parameter) from the gauge parameter $\alpha$ 
\cite{Capri:2016gut}. In order to write the Ward identities that follow from the 
BRST invariance, one adds a set of sources to the action, each coupled to a 
nonlinear BRST variation \cite{Piguet:1995er}, that is
\begin{eqnarray}
S_{ext}&=&\int d^{4}x\bigg(\Omega^{a}_{\mu}\,(sA_\mu^a)
+L^{a}(sc^{a})
+K^{a}\,(s\xi^a)\bigg)\,\nonumber\\
&=&\int d^{4}x\bigg(-\Omega^{a}_{\mu}\,D^{ab}_{\mu}c^{b}
+\frac{g}{2}f^{abc}L^{a}c^{b}c^{c}
+K^{a}\,g^{ab}(\xi)c^{b}\bigg)\,.
\label{eq:external-sources}
\end{eqnarray}

Defining the full classical action as
\begin{eqnarray}\label{eq:class-act}
 \Sigma = S_{RGZ}^{loc} + S_{ext},
\end{eqnarray}
one can express the BRST symmetry in terms of the functional identity \cite{Piguet:1995er} 
\begin{eqnarray}
\mathcal{S}(\Sigma)&\equiv&\int d^{4}x\,\bigg(
\frac{\delta\Sigma}{\delta\Omega^{a}_{\mu}}\frac{\delta\Sigma}{\delta A^{a}_{\mu}}
+\frac{\delta\Sigma}{\delta L^{a}}\frac{\delta\Sigma}{\delta c^{a}}
+\frac{\delta\Sigma}{\delta K^{a}}\frac{\delta\Sigma}{\delta\xi^{a}}
+b^{a}\frac{\delta\Sigma}{\delta\bar{c}^{a}}\bigg)=0,\,
\end{eqnarray}
which is the Slavnov-Taylor identity. 
Furthermore, the equation of motion for the $b$-field,
\begin{equation}
\frac{\delta\Sigma}{\delta b^{a}}=\partial_{\mu}A^{a}_{\mu}-\alpha\,b^{a}\,,
\end{equation}
is regarded as a Ward identity. It implies that the most general counterterm is 
independent of $b^{a}$. 

We also have the identity
\begin{equation}
\frac{\delta\Sigma}{\delta\bar{c}^{a}}+\partial_{\mu}\frac{\delta\Sigma}{\delta\Omega^{a}_{\mu}}=0\,,
\label{eq:anti-ghost}
\end{equation}
which is known as the antighost equation. This identity assures that the field $\bar{c}^{a}$ and the source $\Omega^{a}_{\mu}$ present in the counterterm action appear only in the combination
\begin{equation}
\widehat{\Omega}^{a}_{\mu}=\Omega^{a}_{\mu}+\partial_{\mu}\bar{c}^{a}\,.
\end{equation}

\subsection{Kinematical constraints from Ward identities}

The Ward identities in the previous subsection have been written for the classical 
action (\ref{eq:class-act}).
However, it is well-known that they are also valid for the quantum action $\Gamma$ at all 
orders, since the theory is renormalizable. Therefore, such identities can be used to provide
important relations that correlation functions must obey to all orders of perturbation theory.

Let us now consider two particular cases related to the ghost-gluon vertex, which are known as the Taylor kinematics \cite{Taylor:1971ff}. In the Landau gauge, one can show that
\begin{eqnarray}\label{eq:ghost-WI}
 \int d^4x\,\left(\frac{\delta \Gamma}{\delta c^a(x)} + gf^{abc}\bar c^b
 \frac{\delta\Gamma}{\delta b^c}\right) = g\int d^4x\,f^{abc}\left(\Omega_\mu^b A_\mu^c - L^b c^c\right),
\end{eqnarray}
which is known as the ghost Ward identity. By taking a functional derivative
of (\ref{eq:ghost-WI}) with respect to $A$ and $\Omega$, one finds
\begin{eqnarray}
 \frac{\delta^2}{\delta A_\mu^d(y)\delta\Omega_\nu^e(z)}\int d^4x\,\left(\frac{\delta \Gamma}{\delta c^a(x)} + gf^{abc}\bar c^b
 \frac{\delta\Gamma}{\delta b^c(x)}\right) &=& 
 gf^{abc}\frac{\delta^2}{\delta A_\mu^d(y)\delta\Omega_\nu^e(z)}\int d^4x\left(\Omega_\rho^b A_\rho^c - L^b c^c\right)\nonumber\\
 \int d^4x\,\left(\frac{\delta^3 \Gamma}{\delta A_\mu^d(y)\delta\Omega_\nu^e(z)\delta c^a(x)} + gf^{abc}\bar c^b\frac{\delta^3\Gamma}{\delta A_\mu^d(y)\delta\Omega_\nu^e(z)\delta b^c(x)}\right) &=&
 gf^{abc}\int d^4x\,\delta_{\nu\rho}\delta^{eb}\delta(x-z)\,\delta_{\mu\rho}\delta^{cd}\delta(y-x)\nonumber\\
  \int d^4x\,\left(\frac{\delta^3 \Gamma}{\delta A_\mu^d(y)\delta\Omega_\nu^e(z)\delta c^a(x)}\right) &=&
 gf^{aed}\delta_{\mu\nu}\delta(y-z).
\end{eqnarray}
Now, deriving both sides with respect to $z$ and using the antighost identity 
(\ref{eq:anti-ghost}), one finds
\begin{equation}\label{eq:Taylor-identity-space}
  \partial_\nu^{(z)}\int d^4x\,\left(\frac{\delta^3 \Gamma}{\delta A_\mu^d(y)\delta\Omega_\nu^e(z)\delta c^a(x)}\right) =
 gf^{aed}\delta_{\mu\nu}\partial_\nu^{(z)}\delta(y-z)\,,
 \end{equation}
which implies
 \begin{equation}
 -\int d^4x\,\left(\frac{\delta^3 \Gamma}{\delta A_\mu^d(y)\delta \bar c^e(z)\delta c^a(x)}\right) =
 gf^{aed}\delta_{\mu\nu}\partial_\nu^{(z)}\delta(y-z).
\end{equation}

Taking the Fourier transform of the identity (\ref{eq:Taylor-identity-space}),
we finally see that the ghost-gluon vertex function in the Laudau gauge reduces to
\begin{eqnarray}
 (\Gamma_{A\,\bar c\,c})^{abc}_\mu(-p,p,0) = -i gf^{abc}p_\mu\,,
\end{eqnarray}
at the limit of vanishing ghost momentum, to all orders. 

A second identity can be derived from the antighost equation. Integrating 
(\ref{eq:anti-ghost}), one finds
\begin{equation}
 0 = \int\,d^4x\,\left(\partial_\mu^{(x)}\frac{\delta\Gamma}{\delta\Omega_\mu^a(x)}
 + \frac{\delta\Gamma}{\delta\bar c^a(x)}\right)
 \Longrightarrow \;\;0 = \frac{\delta^2}{\delta A_\mu^d(y)\delta c^e(z)}
 \int\,d^4x\,\left( \frac{\delta\Gamma}{\delta\bar c^a(x)}\right)
 = \int\,d^4x\,\left(\frac{\delta^3\Gamma}{\delta A_\mu^d(y)\delta c^e(z)\delta\bar c^a(x)}\right).
\end{equation}
This means that the vertex function vanishes for zero antighost momentum 
at all orders, that is
\begin{eqnarray}
 (\Gamma_{A\,\bar c\,c})^{abc}_\mu(p,0,-p) = 0.
\end{eqnarray}

Therefore, we find that, thanks to the BRST 
transformations (\ref{npbrst16a}), the nontrivial kinematic 
relations (\ref{eq:Taylor-kin-ghost}) and (\ref{eq:Taylor-kin-antighost}) 
are true not only in Yang-Mills \cite{Taylor:1971ff}, but also 
in the RGZ framework.
In the next appendix, we show the calculation of the $A\bar cc$ vertex 
function within the RGZ framework at one-loop order, verifying that the 
exact results \eqref{eq:Taylor-kin-ghost} and \eqref{eq:Taylor-kin-antighost} 
are indeed satisfied explicitly at this order.

\section{Explicit calculation of the diagrams that contribute to the ghost-gluon vertex}\label{sec:appendix-1-loop-diagrams}

Let us now calculate the one-loop diagrams for the three-point one-particle 
irreducible function 
\begin{eqnarray}\label{eq:1loop-sum-of-diagrams}
 [\Gamma_{A\bar c c} (k,p,-p-k)]_{\mu}^{abc} 
 &=& \vev{A_\mu^a(k)\,\bar c^{b}(p)\,c^{c}(-p-k)}^{1PI}\nonumber\\
 &=& -igf^{abc}p_\mu + (I)^{abc}_\mu+(II)^{abc}_\mu+(III)^{abc}_\mu+(IV)^{abc}_\mu
 + {\cal O}(g^5),
\end{eqnarray}
where each of the four terms above equals the corresponding one-loop diagram 
in Fig. \ref{fig:feyndiags}.
In order to simplify our analysis, we will only 
consider the soft gluon limit ($k\rightarrow 0$).
Since all integrals are IR- and UV-convergent, the $k\rightarrow0$ limit simply amounts to taking $k=0$ in all expressions. 
Let us calculate each of the integrals defined in the previous section in this limit.

\subsection{Diagram I}

Using the Feynman rules listed in Appendix \ref{sec:appendix-feynman-rules},
one finds
\begin{eqnarray}
(I)^{abc}_{\mu} 
&=&i\frac{N g^3}{2}f^{abc} \int_\ell(\ell-p)_\mu\,
D_{AA}(\ell)D_{\bar cc}(\ell-p-k)D_{\bar cc}(\ell-p)\;\left[p\cdot(p+k) - \frac{(p\cdot\ell)[(p+k)\cdot \ell]}{\ell^2}\right].
\end{eqnarray}
The soft-gluon limit of this diagram reads
\begin{eqnarray}\label{eq:diag1-softgluon}
(I)^{abc}_{\mu}(k\rightarrow0)
&=&i\frac{N g^3}{2}f^{abc}\int_\ell(\ell-p)_\mu\,
D_{AA}(\ell)D_{\bar\omega\omega}(\ell-p)D_{\bar\omega\omega}(\ell-p)\;\left[p\cdot p - \frac{(p\cdot\ell)^2]}{\ell^2}\right]\nonumber\\
&=&i\frac{Ng^3}2f^{abc} \left[R_+ J_\mu(a_+;p) + R_- J_\mu(a_-;p)\right].
\end{eqnarray}
where $a_\pm$ are the (complex) poles of the gluon propagator and $R_\pm$ are
their respective residues. Explicitly, 
\begin{eqnarray}
D_{AA}(p^2) &=& \frac{p^2+M^2}{(p^2+m^2)(p^2+M^2)+\Lambda^4} 
\equiv \frac{R_+}{p^2+a_+^2} + \frac{R_-}{p^2+a_-^2},
\end{eqnarray}
with
\begin{eqnarray}
 a_+^2 &=& \frac{m^2+M^2+\sqrt{(m^2-M^2)^2-4\Lambda^4}}{2},\nonumber\\
 a_-^2 &=& \frac{m^2+M^2-\sqrt{(m^2-M^2)^2-4\Lambda^4}}{2},\nonumber\\
 R_+ &=& \frac{m^2-M^2+\sqrt{(m^2-M^2)^2-4\Lambda^4}}{2\sqrt{(m^2-M^2)^2-4\Lambda^4}},\nonumber\\
 R_- &=& \frac{-m^2+M^2-\sqrt{(m^2-M^2)^2-4\Lambda^4}}{2\sqrt{(m^2-M^2)^2-4\Lambda^4}} = 1-R_+,
\end{eqnarray}
where $\Lambda^4 = 2Ng^2\gamma^4$. In (\ref{eq:diag1-softgluon}) 
we used the integral defined by
\begin{equation}\label{eq:J-rho}
 J_\mu(m_1;p):=\int_\ell
\frac{1}{\ell^2}\frac{1}{\ell^2+m_1^2}\frac{p^2\ell^2 - (p\cdot\ell)^2}{[(\ell-p)^2]^2}(\ell-p)_\mu.
\end{equation}

Using the standard technique of Feynman parameters \cite{Peskin:1995ev} 
and integrating in the loop momentum, we find after a straightforward 
calculation,
\begin{eqnarray}
 J_\mu(m_1;p) 
  &=&-\frac{1}{64\pi^2} p_\mu\times\frac{2m_1^2p^2(p^2+m_1^2)+p^6\log\left(1+\frac{m_1^2}{p^2}\right)-(3p^2+2m_1^2)m_1^4\log\left(1+\frac{p^2}{m_1^2}\right)}{m_1^2p^4},
\end{eqnarray}
%

\subsection{Diagram II}

The second diagram of (\ref{eq:1loop-sum-of-diagrams}) is the most complicated 
of the four diagrams, due to the presence of the triple gluon vertex. After 
some simplifications, it is given by
\begin{eqnarray}
(II)^{abc}_{\mu} 
&=&i\frac{Ng^3}{2}f^{abc}p_\phi(k+p)_\gamma\int_\ell D_{\bar cc}(\ell+p)D_{AA}(\ell)D_{AA}(\ell-k)
\mathcal{P}_{\eta\phi}(\ell)\mathcal{P}_{\gamma\delta}(\ell-k)\left[(2\ell-k)_\mu\delta_{\eta\delta}+2k_\eta\delta_{\mu\delta}
-2k_\delta\delta_{\eta\mu}\right].\nonumber\\
\end{eqnarray}
In the soft gluon limit $(k\rightarrow0)$, the tensor structure of diagram 
($II$) simplifies tremendously, and one has
\begin{eqnarray}
 (II)^{abc}_{\mu}(k\rightarrow0) &=&i\frac{Ng^3}{2}f^{abc}p_\phi p_\gamma\int_\ell D_{\bar cc}(\ell+p)D_{AA}(\ell)D_{AA}(\ell)
\mathcal{P}_{\eta\phi}(\ell)\mathcal{P}_{\gamma\delta}(\ell)\left[(2\ell)_\mu\delta_{\eta\delta}\right]\nonumber\\
&=&
iNg^3f^{abc}\bigg[R_+^2K_\mu(a_+,a_+;p) + R_-^2K_\mu(a_-,a_-;p) + 2R_+R_- K_\mu(a_+,a_-;p)\bigg],\nonumber\\
\end{eqnarray}
where we defined the integral
\begin{eqnarray}\label{eq:integral-K}
 K_\mu(m_1,m_2;p)&:=&\int_\ell \frac{1}{(\ell+p)^2}\frac{1}{\ell^2+m_1^2}\frac{1}{\ell^2+m_2^2}
\,\left[\frac{p^2\ell^2 - (p\cdot\ell)^2}{\ell^2}\right]\ell_\mu.
\end{eqnarray}

Once again we use the standard technique to find, for spacetime dimension $d$,
\begin{eqnarray}
 K_\mu(m_1,m_2;p) &=& -\frac{d-1}{2(4\pi)^{d/2}}p^2p_\mu\Gamma(3-d/2)\int_0^1\,dx_1\,dx_2\,dx_3\,dx_4 \,\frac{\delta\left(1-\sum_{i=i}^4x_i\right)\,x_1}{\left[x_1(1-x_1)p^2+x_2m_1^2+x_3m_2^2\right]^{3-d/2}}.
\end{eqnarray}

Therefore, for $d=4$, the integral is finite and yields
\begin{eqnarray}
 K_\mu(m_1,m_2;p) &=& 
 \frac{1}{256\pi^2}p_\mu\frac{1}{m_1^2m_2^2p^4(m_1^2-m_2^2)}\bigg\{2m_1^2m_2^6p^2-2m_1^6m_2^2p^2+3m_1^2m_2^4p^4 - 3m_1^4m_2^2p^4+\nonumber\\
 &&\hspace{0.5cm}+\left.2m_1^8m_2^2\log\left(1+\frac{p^2}{m_1^2}\right)-2m_1^2m_2^8\log\left(1+\frac{p^2}{m_2^2}\right)+ 4m_1^6m_2^2p^2\log\left(1+\frac{p^2}{m_1^2}\right)-\right.\nonumber\\
&&\hspace{0.6cm}\left.- 4m_1^2m_2^6p^2\log\left(1+\frac{p^2}{m_2^2}\right) +
4m_1^2m_2^2p^6\log\left(\frac{p^2+m_2^2}{p^2+m_1^2}\right)+ \right.\nonumber\\
&&\hspace{0.7cm}\left.+
2m_1^2p^8\log\left(1+\frac{m_2^2}{p^2}\right) - 
2m_2^2p^8\log\left(1+\frac{m_1^2}{p^2}\right)\right\}.
\end{eqnarray}

\subsection{Diagram III}

Including the two equal contributions from $\varphi\leftrightarrow\bar\varphi$ in the diagram,
\begin{eqnarray}
(III)^{abc}_{\mu} 
&=& i \frac{Ng^3}4(Ng^2\gamma^4)f^{abc}p_\phi(p+k)_\gamma
\int_\ell\,D_{\bar cc}(\ell)\frac{D_{AA}(\ell-p-k)}{(\ell-p-k)^2+\mu^2}\frac{D_{AA}(\ell-p)}{(\ell-p)^2+\mu^2}
\,\mathcal{P}_{\gamma\delta}(\ell-p-k)\,\mathcal{P}_{\delta\phi}(\ell-p)
(\ell-p)_\mu.\nonumber\\
\end{eqnarray}

Similarly to the diagram $(II)$, the soft-gluon limit of diagram ($III$) 
can also be written in terms of the integrals of the type (\ref{eq:integral-K})
\begin{eqnarray}
(III)^{abc}_{\mu} (k\rightarrow0)
&=& i\frac{Ng^3}4\frac{Ng^2\gamma^4}{\left[a_+^2-a_-^2\right]^2}\;f^{abc}
\left[K_\mu(a_+,a_+;p)+K_\mu(a_-,a_-;p)-2K_\mu(a_+,a_-;p)
\right]
\,.
\end{eqnarray}

\subsection{Diagram IV}

Let us finally consider diagram ($IV$). It is given by
\begin{eqnarray}
(IV)^{abc}_{\mu} 
&=&-ig^3(g\gamma^2)\,f^{aho}f^{hcj}f^{jkn}f^{kmo}\delta^{bd}\delta_{\mu\nu}p_\epsilon(p+k)_\gamma\,k_\phi\,
\int_\ell\,D_{\bar cc}(\ell)\frac{D_{AA}(\ell-p-k)}{(\ell-p-k)^2+\mu^2}D_{AA}(\ell-p)\times\nonumber\\
&&\times\mathcal{P}_{\gamma\delta}(\ell-p-k)\mathcal{P}_{\epsilon\phi}(\ell-p)\,.
\end{eqnarray}

Note that this diagram is proportional to the gluon momentum $k$. Therefore, 
given that the integral is also finite, it does not contribute to the soft-gluon
limit of the vertex function (\ref{eq:1loop-sum-of-diagrams}). 

Finally we would like to point out that all diagrams (I)--(IV) are proportional do $p_\alpha(p+k)_\beta$, 
so that the one-loop contributions to the vertex function 
(which are perfectly finite)
all vanish at the limits $p\rightarrow0$ and $p+k\rightarrow0$, as expected 
from the symmetries of the theory (e.g., the so-called Taylor kinematics 
\cite{Taylor:1971ff}) and discussed in Appendix \ref{sec:appendix-ward}.

\newpage
\bibliography{RGZ-bibliography}

\begin{thebibliography}{49}%
\makeatletter
\providecommand \@ifxundefined [1]{%
 \@ifx{#1\undefined}
}%
\providecommand \@ifnum [1]{%
 \ifnum #1\expandafter \@firstoftwo
 \else \expandafter \@secondoftwo
 \fi
}%
\providecommand \@ifx [1]{%
 \ifx #1\expandafter \@firstoftwo
 \else \expandafter \@secondoftwo
 \fi
}%
\providecommand \natexlab [1]{#1}%
\providecommand \enquote  [1]{``#1''}%
\providecommand \bibnamefont  [1]{#1}%
\providecommand \bibfnamefont [1]{#1}%
\providecommand \citenamefont [1]{#1}%
\providecommand \href@noop [0]{\@secondoftwo}%
\providecommand \href [0]{\begingroup \@sanitize@url \@href}%
\providecommand \@href[1]{\@@startlink{#1}\@@href}%
\providecommand \@@href[1]{\endgroup#1\@@endlink}%
\providecommand \@sanitize@url [0]{\catcode `\\12\catcode `\$12\catcode
  `\&12\catcode `\#12\catcode `\^12\catcode `\_12\catcode `\%12\relax}%
\providecommand \@@startlink[1]{}%
\providecommand \@@endlink[0]{}%
\providecommand \url  [0]{\begingroup\@sanitize@url \@url }%
\providecommand \@url [1]{\endgroup\@href {#1}{\urlprefix }}%
\providecommand \urlprefix  [0]{URL }%
\providecommand \Eprint [0]{\href }%
\providecommand \doibase [0]{http://dx.doi.org/}%
\providecommand \selectlanguage [0]{\@gobble}%
\providecommand \bibinfo  [0]{\@secondoftwo}%
\providecommand \bibfield  [0]{\@secondoftwo}%
\providecommand \translation [1]{[#1]}%
\providecommand \BibitemOpen [0]{}%
\providecommand \bibitemStop [0]{}%
\providecommand \bibitemNoStop [0]{.\EOS\space}%
\providecommand \EOS [0]{\spacefactor3000\relax}%
\providecommand \BibitemShut  [1]{\csname bibitem#1\endcsname}%
\let\auto@bib@innerbib\@empty
\bibitem [{\citenamefont {Berges}\ \emph {et~al.}(2002)\citenamefont {Berges},
  \citenamefont {Tetradis},\ and\ \citenamefont {Wetterich}}]{Berges:2000ew}%
  \BibitemOpen
  \bibfield  {author} {\bibinfo {author} {\bibfnamefont {J.}~\bibnamefont
  {Berges}}, \bibinfo {author} {\bibfnamefont {N.}~\bibnamefont {Tetradis}}, \
  and\ \bibinfo {author} {\bibfnamefont {C.}~\bibnamefont {Wetterich}},\ }\href
  {\doibase 10.1016/S0370-1573(01)00098-9} {\bibfield  {journal} {\bibinfo
  {journal} {Phys. Rept.}\ }\textbf {\bibinfo {volume} {363}},\ \bibinfo
  {pages} {223} (\bibinfo {year} {2002})},\ \Eprint
  {http://arxiv.org/abs/hep-ph/0005122} {arXiv:hep-ph/0005122 [hep-ph]}
  \BibitemShut {NoStop}%
\bibitem [{\citenamefont {Pawlowski}(2007)}]{Pawlowski:2005xe}%
  \BibitemOpen
  \bibfield  {author} {\bibinfo {author} {\bibfnamefont {J.~M.}\ \bibnamefont
  {Pawlowski}},\ }\href {\doibase 10.1016/j.aop.2007.01.007} {\bibfield
  {journal} {\bibinfo  {journal} {Annals Phys.}\ }\textbf {\bibinfo {volume}
  {322}},\ \bibinfo {pages} {2831} (\bibinfo {year} {2007})},\ \Eprint
  {http://arxiv.org/abs/hep-th/0512261} {arXiv:hep-th/0512261 [hep-th]}
  \BibitemShut {NoStop}%
\bibitem [{\citenamefont {Bashir}\ \emph {et~al.}(2012)\citenamefont {Bashir},
  \citenamefont {Chang}, \citenamefont {Cloet}, \citenamefont {El-Bennich},
  \citenamefont {Liu}, \citenamefont {Roberts},\ and\ \citenamefont
  {Tandy}}]{Bashir:2012fs}%
  \BibitemOpen
  \bibfield  {author} {\bibinfo {author} {\bibfnamefont {A.}~\bibnamefont
  {Bashir}}, \bibinfo {author} {\bibfnamefont {L.}~\bibnamefont {Chang}},
  \bibinfo {author} {\bibfnamefont {I.~C.}\ \bibnamefont {Cloet}}, \bibinfo
  {author} {\bibfnamefont {B.}~\bibnamefont {El-Bennich}}, \bibinfo {author}
  {\bibfnamefont {Y.-X.}\ \bibnamefont {Liu}}, \bibinfo {author} {\bibfnamefont
  {C.~D.}\ \bibnamefont {Roberts}}, \ and\ \bibinfo {author} {\bibfnamefont
  {P.~C.}\ \bibnamefont {Tandy}},\ }\href {\doibase 10.1088/0253-6102/58/1/16}
  {\bibfield  {journal} {\bibinfo  {journal} {Commun. Theor. Phys.}\ }\textbf
  {\bibinfo {volume} {58}},\ \bibinfo {pages} {79} (\bibinfo {year} {2012})},\
  \Eprint {http://arxiv.org/abs/1201.3366} {arXiv:1201.3366 [nucl-th]}
  \BibitemShut {NoStop}%
\bibitem [{\citenamefont {Aguilar}\ \emph {et~al.}(2016)\citenamefont
  {Aguilar}, \citenamefont {Binosi},\ and\ \citenamefont
  {Papavassiliou}}]{Aguilar:2015bud}%
  \BibitemOpen
  \bibfield  {author} {\bibinfo {author} {\bibfnamefont {A.~C.}\ \bibnamefont
  {Aguilar}}, \bibinfo {author} {\bibfnamefont {D.}~\bibnamefont {Binosi}}, \
  and\ \bibinfo {author} {\bibfnamefont {J.}~\bibnamefont {Papavassiliou}},\
  }\bibfield  {booktitle} {\emph {\bibinfo {booktitle} {{ECT* Workshop on
  Dyson-Schwinger Equations in Modern Mathematics and Physics (DSEMP2014)
  Trento, Italy, September 22-26, 2014}}},\ }\href {\doibase
  10.1007/s11467-015-0517-6} {\bibfield  {journal} {\bibinfo  {journal} {Front.
  Phys.(Beijing)}\ }\textbf {\bibinfo {volume} {11}},\ \bibinfo {pages}
  {111203} (\bibinfo {year} {2016})},\ \Eprint
  {http://arxiv.org/abs/1511.08361} {arXiv:1511.08361 [hep-ph]} \BibitemShut
  {NoStop}%
\bibitem [{\citenamefont {{Nambu}}\ and\ \citenamefont
  {{Jona-Lasinio}}(1961)}]{Nambu-JonaLasinio1961}%
  \BibitemOpen
  \bibfield  {author} {\bibinfo {author} {\bibfnamefont {Y.}~\bibnamefont
  {{Nambu}}}\ and\ \bibinfo {author} {\bibfnamefont {G.}~\bibnamefont
  {{Jona-Lasinio}}},\ }\href {\doibase 10.1103/PhysRev.122.345} {\bibfield
  {journal} {\bibinfo  {journal} {Phys. Rev.}\ }\textbf {\bibinfo {volume}
  {122}},\ \bibinfo {pages} {345} (\bibinfo {year} {1961})}\BibitemShut
  {NoStop}%
\bibitem [{\citenamefont {Klevansky}(1992)}]{Klevansky:1992qe}%
  \BibitemOpen
  \bibfield  {author} {\bibinfo {author} {\bibfnamefont {S.~P.}\ \bibnamefont
  {Klevansky}},\ }\href {\doibase 10.1103/RevModPhys.64.649} {\bibfield
  {journal} {\bibinfo  {journal} {Rev. Mod. Phys.}\ }\textbf {\bibinfo {volume}
  {64}},\ \bibinfo {pages} {649} (\bibinfo {year} {1992})}\BibitemShut
  {NoStop}%
\bibitem [{\citenamefont {Gell-Mann}\ and\ \citenamefont
  {Levy}(1960)}]{GellMann:1960np}%
  \BibitemOpen
  \bibfield  {author} {\bibinfo {author} {\bibfnamefont {M.}~\bibnamefont
  {Gell-Mann}}\ and\ \bibinfo {author} {\bibfnamefont {M.}~\bibnamefont
  {Levy}},\ }\href {\doibase 10.1007/BF02859738} {\bibfield  {journal}
  {\bibinfo  {journal} {Nuovo Cim.}\ }\textbf {\bibinfo {volume} {16}},\
  \bibinfo {pages} {705} (\bibinfo {year} {1960})}\BibitemShut {NoStop}%
\bibitem [{\citenamefont {{Fukushima}}(2004)}]{Fukushima-PNJL-2003}%
  \BibitemOpen
  \bibfield  {author} {\bibinfo {author} {\bibfnamefont {K.}~\bibnamefont
  {{Fukushima}}},\ }\href {\doibase 10.1016/j.physletb.2004.04.027} {\bibfield
  {journal} {\bibinfo  {journal} {Physics Letters B}\ }\textbf {\bibinfo
  {volume} {591}},\ \bibinfo {pages} {277} (\bibinfo {year} {2004})},\ \Eprint
  {http://arxiv.org/abs/arXiv:hep-ph/0310121} {arXiv:hep-ph/0310121}
  \BibitemShut {NoStop}%
\bibitem [{\citenamefont {Schaefer}\ \emph {et~al.}(2007)\citenamefont
  {Schaefer}, \citenamefont {Pawlowski},\ and\ \citenamefont
  {Wambach}}]{Schaefer:2007pw}%
  \BibitemOpen
  \bibfield  {author} {\bibinfo {author} {\bibfnamefont {B.-J.}\ \bibnamefont
  {Schaefer}}, \bibinfo {author} {\bibfnamefont {J.~M.}\ \bibnamefont
  {Pawlowski}}, \ and\ \bibinfo {author} {\bibfnamefont {J.}~\bibnamefont
  {Wambach}},\ }\href {\doibase 10.1103/PhysRevD.76.074023} {\bibfield
  {journal} {\bibinfo  {journal} {Phys. Rev.}\ }\textbf {\bibinfo {volume}
  {D76}},\ \bibinfo {pages} {074023} (\bibinfo {year} {2007})},\ \Eprint
  {http://arxiv.org/abs/0704.3234} {arXiv:0704.3234 [hep-ph]} \BibitemShut
  {NoStop}%
\bibitem [{\citenamefont {Gribov}(1978)}]{Gribov:1977wm}%
  \BibitemOpen
  \bibfield  {author} {\bibinfo {author} {\bibfnamefont {V.~N.}\ \bibnamefont
  {Gribov}},\ }\href {\doibase 10.1016/0550-3213(78)90175-X} {\bibfield
  {journal} {\bibinfo  {journal} {Nucl. Phys.}\ }\textbf {\bibinfo {volume}
  {B139}},\ \bibinfo {pages} {1} (\bibinfo {year} {1978})}\BibitemShut
  {NoStop}%
\bibitem [{Note1()}]{Note1}%
  \BibitemOpen
  \bibinfo {note} {The boundary of the Gribov region is called the Gribov
  horizon.}\BibitemShut {Stop}%
\bibitem [{\citenamefont {Zwanziger}(1989)}]{Zwanziger:1989mf}%
  \BibitemOpen
  \bibfield  {author} {\bibinfo {author} {\bibfnamefont {D.}~\bibnamefont
  {Zwanziger}},\ }\href {\doibase 10.1016/0550-3213(89)90122-3} {\bibfield
  {journal} {\bibinfo  {journal} {Nucl. Phys.}\ }\textbf {\bibinfo {volume}
  {B323}},\ \bibinfo {pages} {513} (\bibinfo {year} {1989})}\BibitemShut
  {NoStop}%
\bibitem [{\citenamefont {Vandersickel}\ and\ \citenamefont
  {Zwanziger}(2012)}]{Vandersickel:2012tz}%
  \BibitemOpen
  \bibfield  {author} {\bibinfo {author} {\bibfnamefont {N.}~\bibnamefont
  {Vandersickel}}\ and\ \bibinfo {author} {\bibfnamefont {D.}~\bibnamefont
  {Zwanziger}},\ }\href {\doibase 10.1016/j.physrep.2012.07.003} {\bibfield
  {journal} {\bibinfo  {journal} {Phys. Rept.}\ }\textbf {\bibinfo {volume}
  {520}},\ \bibinfo {pages} {175} (\bibinfo {year} {2012})},\ \Eprint
  {http://arxiv.org/abs/1202.1491} {arXiv:1202.1491 [hep-th]} \BibitemShut
  {NoStop}%
\bibitem [{\citenamefont {Vandersickel}(2011)}]{Vandersickel:2011zc}%
  \BibitemOpen
  \bibfield  {author} {\bibinfo {author} {\bibfnamefont {N.}~\bibnamefont
  {Vandersickel}},\ }\emph {\bibinfo {title} {{A Study of the Gribov-Zwanziger
  action: from propagators to glueballs}}},\ \href
  {http://inspirehep.net/record/895203/files/arXiv:1104.1315.pdf} {Ph.D.
  thesis},\ \bibinfo  {school} {Gent U.} (\bibinfo {year} {2011}),\ \Eprint
  {http://arxiv.org/abs/1104.1315} {arXiv:1104.1315 [hep-th]} \BibitemShut
  {NoStop}%
\bibitem [{\citenamefont {Sobreiro}\ and\ \citenamefont
  {Sorella}(2005)}]{Sobreiro:2005ec}%
  \BibitemOpen
  \bibfield  {author} {\bibinfo {author} {\bibfnamefont {R.~F.}\ \bibnamefont
  {Sobreiro}}\ and\ \bibinfo {author} {\bibfnamefont {S.~P.}\ \bibnamefont
  {Sorella}},\ }in\ \href@noop {} {\emph {\bibinfo {booktitle} {{13th Jorge
  Andre Swieca Summer School on Particle and Fields Campos do Jordao, Brazil,
  January 9-22, 2005}}}}\ (\bibinfo {year} {2005})\ \Eprint
  {http://arxiv.org/abs/hep-th/0504095} {arXiv:hep-th/0504095 [hep-th]}
  \BibitemShut {NoStop}%
\bibitem [{\citenamefont {Dudal}\ \emph
  {et~al.}(2008{\natexlab{a}})\citenamefont {Dudal}, \citenamefont {Gracey},
  \citenamefont {Sorella}, \citenamefont {Vandersickel},\ and\ \citenamefont
  {Verschelde}}]{Dudal:2008sp}%
  \BibitemOpen
  \bibfield  {author} {\bibinfo {author} {\bibfnamefont {D.}~\bibnamefont
  {Dudal}}, \bibinfo {author} {\bibfnamefont {J.~A.}\ \bibnamefont {Gracey}},
  \bibinfo {author} {\bibfnamefont {S.~P.}\ \bibnamefont {Sorella}}, \bibinfo
  {author} {\bibfnamefont {N.}~\bibnamefont {Vandersickel}}, \ and\ \bibinfo
  {author} {\bibfnamefont {H.}~\bibnamefont {Verschelde}},\ }\href {\doibase
  10.1103/PhysRevD.78.065047} {\bibfield  {journal} {\bibinfo  {journal} {Phys.
  Rev.}\ }\textbf {\bibinfo {volume} {D78}},\ \bibinfo {pages} {065047}
  (\bibinfo {year} {2008}{\natexlab{a}})},\ \Eprint
  {http://arxiv.org/abs/0806.4348} {arXiv:0806.4348 [hep-th]} \BibitemShut
  {NoStop}%
\bibitem [{\citenamefont {Dudal}\ \emph {et~al.}(2010)\citenamefont {Dudal},
  \citenamefont {Oliveira},\ and\ \citenamefont {Vandersickel}}]{Dudal:2010tf}%
  \BibitemOpen
  \bibfield  {author} {\bibinfo {author} {\bibfnamefont {D.}~\bibnamefont
  {Dudal}}, \bibinfo {author} {\bibfnamefont {O.}~\bibnamefont {Oliveira}}, \
  and\ \bibinfo {author} {\bibfnamefont {N.}~\bibnamefont {Vandersickel}},\
  }\href {\doibase 10.1103/PhysRevD.81.074505} {\bibfield  {journal} {\bibinfo
  {journal} {Phys. Rev.}\ }\textbf {\bibinfo {volume} {D81}},\ \bibinfo {pages}
  {074505} (\bibinfo {year} {2010})},\ \Eprint {http://arxiv.org/abs/1002.2374}
  {arXiv:1002.2374 [hep-lat]} \BibitemShut {NoStop}%
\bibitem [{\citenamefont {Capri}\ \emph {et~al.}(2015)\citenamefont {Capri},
  \citenamefont {Dudal}, \citenamefont {Fiorentini}, \citenamefont {Guimaraes},
  \citenamefont {Justo}, \citenamefont {Pereira}, \citenamefont {Mintz},
  \citenamefont {Palhares}, \citenamefont {Sobreiro},\ and\ \citenamefont
  {Sorella}}]{Capri:2015ixa}%
  \BibitemOpen
  \bibfield  {author} {\bibinfo {author} {\bibfnamefont {M.~A.~L.}\
  \bibnamefont {Capri}}, \bibinfo {author} {\bibfnamefont {D.}~\bibnamefont
  {Dudal}}, \bibinfo {author} {\bibfnamefont {D.}~\bibnamefont {Fiorentini}},
  \bibinfo {author} {\bibfnamefont {M.~S.}\ \bibnamefont {Guimaraes}}, \bibinfo
  {author} {\bibfnamefont {I.~F.}\ \bibnamefont {Justo}}, \bibinfo {author}
  {\bibfnamefont {A.~D.}\ \bibnamefont {Pereira}}, \bibinfo {author}
  {\bibfnamefont {B.~W.}\ \bibnamefont {Mintz}}, \bibinfo {author}
  {\bibfnamefont {L.~F.}\ \bibnamefont {Palhares}}, \bibinfo {author}
  {\bibfnamefont {R.~F.}\ \bibnamefont {Sobreiro}}, \ and\ \bibinfo {author}
  {\bibfnamefont {S.~P.}\ \bibnamefont {Sorella}},\ }\href {\doibase
  10.1103/PhysRevD.92.045039} {\bibfield  {journal} {\bibinfo  {journal} {Phys.
  Rev.}\ }\textbf {\bibinfo {volume} {D92}},\ \bibinfo {pages} {045039}
  (\bibinfo {year} {2015})},\ \Eprint {http://arxiv.org/abs/1506.06995}
  {arXiv:1506.06995 [hep-th]} \BibitemShut {NoStop}%
\bibitem [{\citenamefont {Capri}\ \emph
  {et~al.}(2016{\natexlab{a}})\citenamefont {Capri}, \citenamefont
  {Fiorentini}, \citenamefont {Guimaraes}, \citenamefont {Mintz}, \citenamefont
  {Palhares}, \citenamefont {Sorella}, \citenamefont {Dudal}, \citenamefont
  {Justo}, \citenamefont {Pereira},\ and\ \citenamefont
  {Sobreiro}}]{Capri:2015nzw}%
  \BibitemOpen
  \bibfield  {author} {\bibinfo {author} {\bibfnamefont {M.~A.~L.}\
  \bibnamefont {Capri}}, \bibinfo {author} {\bibfnamefont {D.}~\bibnamefont
  {Fiorentini}}, \bibinfo {author} {\bibfnamefont {M.~S.}\ \bibnamefont
  {Guimaraes}}, \bibinfo {author} {\bibfnamefont {B.~W.}\ \bibnamefont
  {Mintz}}, \bibinfo {author} {\bibfnamefont {L.~F.}\ \bibnamefont {Palhares}},
  \bibinfo {author} {\bibfnamefont {S.~P.}\ \bibnamefont {Sorella}}, \bibinfo
  {author} {\bibfnamefont {D.}~\bibnamefont {Dudal}}, \bibinfo {author}
  {\bibfnamefont {I.~F.}\ \bibnamefont {Justo}}, \bibinfo {author}
  {\bibfnamefont {A.~D.}\ \bibnamefont {Pereira}}, \ and\ \bibinfo {author}
  {\bibfnamefont {R.~F.}\ \bibnamefont {Sobreiro}},\ }\href {\doibase
  10.1103/PhysRevD.93.065019} {\bibfield  {journal} {\bibinfo  {journal} {Phys.
  Rev.}\ }\textbf {\bibinfo {volume} {D93}},\ \bibinfo {pages} {065019}
  (\bibinfo {year} {2016}{\natexlab{a}})},\ \Eprint
  {http://arxiv.org/abs/1512.05833} {arXiv:1512.05833 [hep-th]} \BibitemShut
  {NoStop}%
\bibitem [{\citenamefont {Capri}\ \emph
  {et~al.}(2016{\natexlab{b}})\citenamefont {Capri}, \citenamefont {Dudal},
  \citenamefont {Fiorentini}, \citenamefont {Guimaraes}, \citenamefont {Justo},
  \citenamefont {Pereira}, \citenamefont {Mintz}, \citenamefont {Palhares},
  \citenamefont {Sobreiro},\ and\ \citenamefont {Sorella}}]{Capri:2016aqq}%
  \BibitemOpen
  \bibfield  {author} {\bibinfo {author} {\bibfnamefont {M.~A.~L.}\
  \bibnamefont {Capri}}, \bibinfo {author} {\bibfnamefont {D.}~\bibnamefont
  {Dudal}}, \bibinfo {author} {\bibfnamefont {D.}~\bibnamefont {Fiorentini}},
  \bibinfo {author} {\bibfnamefont {M.~S.}\ \bibnamefont {Guimaraes}}, \bibinfo
  {author} {\bibfnamefont {I.~F.}\ \bibnamefont {Justo}}, \bibinfo {author}
  {\bibfnamefont {A.~D.}\ \bibnamefont {Pereira}}, \bibinfo {author}
  {\bibfnamefont {B.~W.}\ \bibnamefont {Mintz}}, \bibinfo {author}
  {\bibfnamefont {L.~F.}\ \bibnamefont {Palhares}}, \bibinfo {author}
  {\bibfnamefont {R.~F.}\ \bibnamefont {Sobreiro}}, \ and\ \bibinfo {author}
  {\bibfnamefont {S.~P.}\ \bibnamefont {Sorella}},\ }\href {\doibase
  10.1103/PhysRevD.94.025035} {\bibfield  {journal} {\bibinfo  {journal} {Phys.
  Rev.}\ }\textbf {\bibinfo {volume} {D94}},\ \bibinfo {pages} {025035}
  (\bibinfo {year} {2016}{\natexlab{b}})},\ \Eprint
  {http://arxiv.org/abs/1605.02610} {arXiv:1605.02610 [hep-th]} \BibitemShut
  {NoStop}%
\bibitem [{\citenamefont {Gracey}(2012)}]{Gracey:2012wf}%
  \BibitemOpen
  \bibfield  {author} {\bibinfo {author} {\bibfnamefont {J.~A.}\ \bibnamefont
  {Gracey}},\ }\href {\doibase 10.1103/PhysRevD.86.105029} {\bibfield
  {journal} {\bibinfo  {journal} {Phys. Rev.}\ }\textbf {\bibinfo {volume}
  {D86}},\ \bibinfo {pages} {105029} (\bibinfo {year} {2012})},\ \Eprint
  {http://arxiv.org/abs/1210.5962} {arXiv:1210.5962 [hep-th]} \BibitemShut
  {NoStop}%
\bibitem [{\citenamefont {Capri}\ \emph
  {et~al.}(2016{\natexlab{c}})\citenamefont {Capri}, \citenamefont
  {Fiorentini}, \citenamefont {Guimaraes}, \citenamefont {Mintz}, \citenamefont
  {Palhares},\ and\ \citenamefont {Sorella}}]{Fiorentini:2016rwx}%
  \BibitemOpen
  \bibfield  {author} {\bibinfo {author} {\bibfnamefont {M.~A.~L.}\
  \bibnamefont {Capri}}, \bibinfo {author} {\bibfnamefont {D.}~\bibnamefont
  {Fiorentini}}, \bibinfo {author} {\bibfnamefont {M.~S.}\ \bibnamefont
  {Guimaraes}}, \bibinfo {author} {\bibfnamefont {B.~W.}\ \bibnamefont
  {Mintz}}, \bibinfo {author} {\bibfnamefont {L.~F.}\ \bibnamefont {Palhares}},
  \ and\ \bibinfo {author} {\bibfnamefont {S.~P.}\ \bibnamefont {Sorella}},\
  }\href {\doibase 10.1103/PhysRevD.94.065009} {\bibfield  {journal} {\bibinfo
  {journal} {Phys. Rev.}\ }\textbf {\bibinfo {volume} {D94}},\ \bibinfo {pages}
  {065009} (\bibinfo {year} {2016}{\natexlab{c}})},\ \Eprint
  {http://arxiv.org/abs/1606.06601} {arXiv:1606.06601 [hep-th]} \BibitemShut
  {NoStop}%
\bibitem [{\citenamefont {Lavelle}\ and\ \citenamefont
  {McMullan}(1997)}]{Lavelle:1995ty}%
  \BibitemOpen
  \bibfield  {author} {\bibinfo {author} {\bibfnamefont {M.}~\bibnamefont
  {Lavelle}}\ and\ \bibinfo {author} {\bibfnamefont {D.}~\bibnamefont
  {McMullan}},\ }\href {\doibase 10.1016/S0370-1573(96)00019-1} {\bibfield
  {journal} {\bibinfo  {journal} {Phys. Rept.}\ }\textbf {\bibinfo {volume}
  {279}},\ \bibinfo {pages} {1} (\bibinfo {year} {1997})},\ \Eprint
  {http://arxiv.org/abs/hep-ph/9509344} {arXiv:hep-ph/9509344 [hep-ph]}
  \BibitemShut {NoStop}%
\bibitem [{\citenamefont {Capri}\ \emph
  {et~al.}(2017{\natexlab{a}})\citenamefont {Capri}, \citenamefont {Dudal},
  \citenamefont {Pereira}, \citenamefont {Fiorentini}, \citenamefont
  {Guimaraes}, \citenamefont {Mintz}, \citenamefont {Palhares},\ and\
  \citenamefont {Sorella}}]{Capri:2016gut}%
  \BibitemOpen
  \bibfield  {author} {\bibinfo {author} {\bibfnamefont {M.~A.~L.}\
  \bibnamefont {Capri}}, \bibinfo {author} {\bibfnamefont {D.}~\bibnamefont
  {Dudal}}, \bibinfo {author} {\bibfnamefont {A.~D.}\ \bibnamefont {Pereira}},
  \bibinfo {author} {\bibfnamefont {D.}~\bibnamefont {Fiorentini}}, \bibinfo
  {author} {\bibfnamefont {M.~S.}\ \bibnamefont {Guimaraes}}, \bibinfo {author}
  {\bibfnamefont {B.~W.}\ \bibnamefont {Mintz}}, \bibinfo {author}
  {\bibfnamefont {L.~F.}\ \bibnamefont {Palhares}}, \ and\ \bibinfo {author}
  {\bibfnamefont {S.~P.}\ \bibnamefont {Sorella}},\ }\href {\doibase
  10.1103/PhysRevD.95.045011} {\bibfield  {journal} {\bibinfo  {journal} {Phys.
  Rev.}\ }\textbf {\bibinfo {volume} {D95}},\ \bibinfo {pages} {045011}
  (\bibinfo {year} {2017}{\natexlab{a}})},\ \Eprint
  {http://arxiv.org/abs/1611.10077} {arXiv:1611.10077 [hep-th]} \BibitemShut
  {NoStop}%
\bibitem [{Note2()}]{Note2}%
  \BibitemOpen
  \bibinfo {note} {We thank M. Tissier, N. Wschebor and U. Reinosa for having
  called our attention to this important point.}\BibitemShut {Stop}%
\bibitem [{\citenamefont {Ferrari}\ and\ \citenamefont
  {Quadri}(2004)}]{Ferrari:2004pd}%
  \BibitemOpen
  \bibfield  {author} {\bibinfo {author} {\bibfnamefont {R.}~\bibnamefont
  {Ferrari}}\ and\ \bibinfo {author} {\bibfnamefont {A.}~\bibnamefont
  {Quadri}},\ }\href {\doibase 10.1088/1126-6708/2004/11/019} {\bibfield
  {journal} {\bibinfo  {journal} {JHEP}\ }\textbf {\bibinfo {volume} {11}},\
  \bibinfo {pages} {019} (\bibinfo {year} {2004})},\ \Eprint
  {http://arxiv.org/abs/hep-th/0408168} {arXiv:hep-th/0408168 [hep-th]}
  \BibitemShut {NoStop}%
\bibitem [{\citenamefont {Capri}\ \emph
  {et~al.}(2017{\natexlab{b}})\citenamefont {Capri}, \citenamefont
  {Fiorentini}, \citenamefont {Pereira},\ and\ \citenamefont
  {Sorella}}]{Capri:2017bfd}%
  \BibitemOpen
  \bibfield  {author} {\bibinfo {author} {\bibfnamefont {M.~A.~L.}\
  \bibnamefont {Capri}}, \bibinfo {author} {\bibfnamefont {D.}~\bibnamefont
  {Fiorentini}}, \bibinfo {author} {\bibfnamefont {A.~D.}\ \bibnamefont
  {Pereira}}, \ and\ \bibinfo {author} {\bibfnamefont {S.~P.}\ \bibnamefont
  {Sorella}},\ }\href {\doibase 10.1103/PhysRevD.96.054022} {\bibfield
  {journal} {\bibinfo  {journal} {Phys. Rev.}\ }\textbf {\bibinfo {volume}
  {D96}},\ \bibinfo {pages} {054022} (\bibinfo {year} {2017}{\natexlab{b}})},\
  \Eprint {http://arxiv.org/abs/1708.01543} {arXiv:1708.01543 [hep-th]}
  \BibitemShut {NoStop}%
\bibitem [{\citenamefont {Dudal}\ \emph
  {et~al.}(2008{\natexlab{b}})\citenamefont {Dudal}, \citenamefont {Sorella},
  \citenamefont {Vandersickel},\ and\ \citenamefont
  {Verschelde}}]{Dudal:2007cw}%
  \BibitemOpen
  \bibfield  {author} {\bibinfo {author} {\bibfnamefont {D.}~\bibnamefont
  {Dudal}}, \bibinfo {author} {\bibfnamefont {S.~P.}\ \bibnamefont {Sorella}},
  \bibinfo {author} {\bibfnamefont {N.}~\bibnamefont {Vandersickel}}, \ and\
  \bibinfo {author} {\bibfnamefont {H.}~\bibnamefont {Verschelde}},\ }\href
  {\doibase 10.1103/PhysRevD.77.071501} {\bibfield  {journal} {\bibinfo
  {journal} {Phys. Rev.}\ }\textbf {\bibinfo {volume} {D77}},\ \bibinfo {pages}
  {071501} (\bibinfo {year} {2008}{\natexlab{b}})},\ \Eprint
  {http://arxiv.org/abs/0711.4496} {arXiv:0711.4496 [hep-th]} \BibitemShut
  {NoStop}%
\bibitem [{\citenamefont {Gracey}(2010)}]{Gracey:2010cg}%
  \BibitemOpen
  \bibfield  {author} {\bibinfo {author} {\bibfnamefont {J.~A.}\ \bibnamefont
  {Gracey}},\ }\href {\doibase 10.1103/PhysRevD.82.085032} {\bibfield
  {journal} {\bibinfo  {journal} {Phys. Rev.}\ }\textbf {\bibinfo {volume}
  {D82}},\ \bibinfo {pages} {085032} (\bibinfo {year} {2010})},\ \Eprint
  {http://arxiv.org/abs/1009.3889} {arXiv:1009.3889 [hep-th]} \BibitemShut
  {NoStop}%
\bibitem [{\citenamefont {Dudal}\ \emph {et~al.}(2011)\citenamefont {Dudal},
  \citenamefont {Sorella},\ and\ \citenamefont {Vandersickel}}]{Dudal:2011gd}%
  \BibitemOpen
  \bibfield  {author} {\bibinfo {author} {\bibfnamefont {D.}~\bibnamefont
  {Dudal}}, \bibinfo {author} {\bibfnamefont {S.~P.}\ \bibnamefont {Sorella}},
  \ and\ \bibinfo {author} {\bibfnamefont {N.}~\bibnamefont {Vandersickel}},\
  }\href {\doibase 10.1103/PhysRevD.84.065039} {\bibfield  {journal} {\bibinfo
  {journal} {Phys. Rev.}\ }\textbf {\bibinfo {volume} {D84}},\ \bibinfo {pages}
  {065039} (\bibinfo {year} {2011})},\ \Eprint {http://arxiv.org/abs/1105.3371}
  {arXiv:1105.3371 [hep-th]} \BibitemShut {NoStop}%
\bibitem [{Note3()}]{Note3}%
  \BibitemOpen
  \bibinfo {note} {One should note that $\left [\protect \EuScript {M}(A^h)+\mu
  ^2\right ]^{ab}$ is a shorthand notation for $\protect \EuScript
  {M}^{ab}(A^h)+\mu ^2\delta ^{ab}$.}\BibitemShut {Stop}%
\bibitem [{Note4()}]{Note4}%
  \BibitemOpen
  \bibinfo {note} {For the complete set of BRST transformations of the fields
  in the action (\ref {eq:RGZ-local-action}) we refer to \cite
  {Capri:2017bfd}.}\BibitemShut {Stop}%
\bibitem [{Note5()}]{Note5}%
  \BibitemOpen
  \bibinfo {note} {Another equivalent possible formulation of the theory would
  include nonlocal, momentum-dependent vertices instead of the auxiliary
  fields. However, for the sake of using standard QFT techniques, we employ the
  local version of the theory.}\BibitemShut {Stop}%
\bibitem [{\citenamefont {Ball}\ and\ \citenamefont
  {Chiu}(1980)}]{Ball:1980ax}%
  \BibitemOpen
  \bibfield  {author} {\bibinfo {author} {\bibfnamefont {J.~S.}\ \bibnamefont
  {Ball}}\ and\ \bibinfo {author} {\bibfnamefont {T.-W.}\ \bibnamefont
  {Chiu}},\ }\href {\doibase 10.1103/PhysRevD.22.2550,
  10.1103/PhysRevD.23.3085} {\bibfield  {journal} {\bibinfo  {journal} {Phys.
  Rev.}\ }\textbf {\bibinfo {volume} {D22}},\ \bibinfo {pages} {2550} (\bibinfo
  {year} {1980})},\ \bibinfo {note} {[Erratum: Phys.
  Rev.D23,3085(1981)]}\BibitemShut {NoStop}%
\bibitem [{\citenamefont {Taylor}(1971)}]{Taylor:1971ff}%
  \BibitemOpen
  \bibfield  {author} {\bibinfo {author} {\bibfnamefont {J.~C.}\ \bibnamefont
  {Taylor}},\ }\href {\doibase 10.1016/0550-3213(71)90297-5} {\bibfield
  {journal} {\bibinfo  {journal} {Nucl. Phys.}\ }\textbf {\bibinfo {volume}
  {B33}},\ \bibinfo {pages} {436} (\bibinfo {year} {1971})}\BibitemShut
  {NoStop}%
\bibitem [{Note6()}]{Note6}%
  \BibitemOpen
  \bibinfo {note} {For large momenta, perturbative logarithmic corrections must
  be added to this tree-level expression in order to have good agreement with
  lattice data.}\BibitemShut {Stop}%
\bibitem [{\citenamefont {Cucchieri}\ \emph {et~al.}(2012)\citenamefont
  {Cucchieri}, \citenamefont {Dudal}, \citenamefont {Mendes},\ and\
  \citenamefont {Vandersickel}}]{Cucchieri:2011ig}%
  \BibitemOpen
  \bibfield  {author} {\bibinfo {author} {\bibfnamefont {A.}~\bibnamefont
  {Cucchieri}}, \bibinfo {author} {\bibfnamefont {D.}~\bibnamefont {Dudal}},
  \bibinfo {author} {\bibfnamefont {T.}~\bibnamefont {Mendes}}, \ and\ \bibinfo
  {author} {\bibfnamefont {N.}~\bibnamefont {Vandersickel}},\ }\href {\doibase
  10.1103/PhysRevD.85.094513} {\bibfield  {journal} {\bibinfo  {journal} {Phys.
  Rev.}\ }\textbf {\bibinfo {volume} {D85}},\ \bibinfo {pages} {094513}
  (\bibinfo {year} {2012})},\ \Eprint {http://arxiv.org/abs/1111.2327}
  {arXiv:1111.2327 [hep-lat]} \BibitemShut {NoStop}%
\bibitem [{\citenamefont {Oliveira}\ and\ \citenamefont
  {Silva}(2012)}]{Oliveira:2012eh}%
  \BibitemOpen
  \bibfield  {author} {\bibinfo {author} {\bibfnamefont {O.}~\bibnamefont
  {Oliveira}}\ and\ \bibinfo {author} {\bibfnamefont {P.~J.}\ \bibnamefont
  {Silva}},\ }\href {\doibase 10.1103/PhysRevD.86.114513} {\bibfield  {journal}
  {\bibinfo  {journal} {Phys. Rev.}\ }\textbf {\bibinfo {volume} {D86}},\
  \bibinfo {pages} {114513} (\bibinfo {year} {2012})},\ \Eprint
  {http://arxiv.org/abs/1207.3029} {arXiv:1207.3029 [hep-lat]} \BibitemShut
  {NoStop}%
\bibitem [{Note7()}]{Note7}%
  \BibitemOpen
  \bibinfo {note} {Note that the fits presented in the cited lattice references
  include an overall renormalization factor as an extra fit parameter. In our
  RGZ model, this factor is fixed at $Z=1$ at this order of perturbation
  theory. Nevertheless, the fixing of this overall constant can be seen as a
  renormalization condition for the propagator.}\BibitemShut {Stop}%
\bibitem [{\citenamefont {Peskin}\ and\ \citenamefont
  {Schroeder}(1995)}]{Peskin:1995ev}%
  \BibitemOpen
  \bibfield  {author} {\bibinfo {author} {\bibfnamefont {M.~E.}\ \bibnamefont
  {Peskin}}\ and\ \bibinfo {author} {\bibfnamefont {D.~V.}\ \bibnamefont
  {Schroeder}},\ }\href {http://www.slac.stanford.edu/~mpeskin/QFT.html} {\emph
  {\bibinfo {title} {{An Introduction to quantum field theory}}}}\ (\bibinfo
  {publisher} {Addison-Wesley},\ \bibinfo {address} {Reading, USA},\ \bibinfo
  {year} {1995})\BibitemShut {NoStop}%
\bibitem [{Note8()}]{Note8}%
  \BibitemOpen
  \bibinfo {note} {Due to large error bars in lattice data, it is not possible
  to exclude a crossing below the tree level value for low
  momenta.}\BibitemShut {Stop}%
\bibitem [{\citenamefont {Cyrol}\ \emph {et~al.}(2016)\citenamefont {Cyrol},
  \citenamefont {Fister}, \citenamefont {Mitter}, \citenamefont {Pawlowski},\
  and\ \citenamefont {Strodthoff}}]{Cyrol:2016tym}%
  \BibitemOpen
  \bibfield  {author} {\bibinfo {author} {\bibfnamefont {A.~K.}\ \bibnamefont
  {Cyrol}}, \bibinfo {author} {\bibfnamefont {L.}~\bibnamefont {Fister}},
  \bibinfo {author} {\bibfnamefont {M.}~\bibnamefont {Mitter}}, \bibinfo
  {author} {\bibfnamefont {J.~M.}\ \bibnamefont {Pawlowski}}, \ and\ \bibinfo
  {author} {\bibfnamefont {N.}~\bibnamefont {Strodthoff}},\ }\href {\doibase
  10.1103/PhysRevD.94.054005} {\bibfield  {journal} {\bibinfo  {journal} {Phys.
  Rev.}\ }\textbf {\bibinfo {volume} {D94}},\ \bibinfo {pages} {054005}
  (\bibinfo {year} {2016})},\ \Eprint {http://arxiv.org/abs/1605.01856}
  {arXiv:1605.01856 [hep-ph]} \BibitemShut {NoStop}%
\bibitem [{\citenamefont {Cucchieri}\ \emph {et~al.}(2008)\citenamefont
  {Cucchieri}, \citenamefont {Maas},\ and\ \citenamefont
  {Mendes}}]{Cucchieri:2008qm}%
  \BibitemOpen
  \bibfield  {author} {\bibinfo {author} {\bibfnamefont {A.}~\bibnamefont
  {Cucchieri}}, \bibinfo {author} {\bibfnamefont {A.}~\bibnamefont {Maas}}, \
  and\ \bibinfo {author} {\bibfnamefont {T.}~\bibnamefont {Mendes}},\ }\href
  {\doibase 10.1103/PhysRevD.77.094510} {\bibfield  {journal} {\bibinfo
  {journal} {Phys. Rev.}\ }\textbf {\bibinfo {volume} {D77}},\ \bibinfo {pages}
  {094510} (\bibinfo {year} {2008})},\ \Eprint {http://arxiv.org/abs/0803.1798}
  {arXiv:0803.1798 [hep-lat]} \BibitemShut {NoStop}%
\bibitem [{\citenamefont {Pelaez}\ \emph {et~al.}(2013)\citenamefont {Pelaez},
  \citenamefont {Tissier},\ and\ \citenamefont {Wschebor}}]{Pelaez:2013cpa}%
  \BibitemOpen
  \bibfield  {author} {\bibinfo {author} {\bibfnamefont {M.}~\bibnamefont
  {Pelaez}}, \bibinfo {author} {\bibfnamefont {M.}~\bibnamefont {Tissier}}, \
  and\ \bibinfo {author} {\bibfnamefont {N.}~\bibnamefont {Wschebor}},\ }\href
  {\doibase 10.1103/PhysRevD.88.125003} {\bibfield  {journal} {\bibinfo
  {journal} {Phys. Rev.}\ }\textbf {\bibinfo {volume} {D88}},\ \bibinfo {pages}
  {125003} (\bibinfo {year} {2013})},\ \Eprint {http://arxiv.org/abs/1310.2594}
  {arXiv:1310.2594 [hep-th]} \BibitemShut {NoStop}%
\bibitem [{\citenamefont {Tissier}\ and\ \citenamefont
  {Wschebor}(2011)}]{Tissier:2011ey}%
  \BibitemOpen
  \bibfield  {author} {\bibinfo {author} {\bibfnamefont {M.}~\bibnamefont
  {Tissier}}\ and\ \bibinfo {author} {\bibfnamefont {N.}~\bibnamefont
  {Wschebor}},\ }\href {\doibase 10.1103/PhysRevD.84.045018} {\bibfield
  {journal} {\bibinfo  {journal} {Phys. Rev.}\ }\textbf {\bibinfo {volume}
  {D84}},\ \bibinfo {pages} {045018} (\bibinfo {year} {2011})},\ \Eprint
  {http://arxiv.org/abs/1105.2475} {arXiv:1105.2475 [hep-th]} \BibitemShut
  {NoStop}%
\bibitem [{\citenamefont {Ilgenfritz}\ \emph {et~al.}(2007)\citenamefont
  {Ilgenfritz}, \citenamefont {Muller-Preussker}, \citenamefont {Sternbeck},
  \citenamefont {Schiller},\ and\ \citenamefont
  {Bogolubsky}}]{Ilgenfritz:2006he}%
  \BibitemOpen
  \bibfield  {author} {\bibinfo {author} {\bibfnamefont {E.~M.}\ \bibnamefont
  {Ilgenfritz}}, \bibinfo {author} {\bibfnamefont {M.}~\bibnamefont
  {Muller-Preussker}}, \bibinfo {author} {\bibfnamefont {A.}~\bibnamefont
  {Sternbeck}}, \bibinfo {author} {\bibfnamefont {A.}~\bibnamefont {Schiller}},
  \ and\ \bibinfo {author} {\bibfnamefont {I.~L.}\ \bibnamefont {Bogolubsky}},\
  }\bibfield  {booktitle} {\emph {\bibinfo {booktitle} {{Infrared QCD in Rio:
  Propagators, condensates and topological effects. Proceedings, International
  Meeting, IRQCD 2006, Rio de Janeiro, Brazil, June 5-9, 2006}}},\ }\href
  {\doibase 10.1590/S0103-97332007000200006} {\bibfield  {journal} {\bibinfo
  {journal} {Braz. J. Phys.}\ }\textbf {\bibinfo {volume} {37}},\ \bibinfo
  {pages} {193} (\bibinfo {year} {2007})},\ \Eprint
  {http://arxiv.org/abs/hep-lat/0609043} {arXiv:hep-lat/0609043 [hep-lat]}
  \BibitemShut {NoStop}%
\bibitem [{\citenamefont {Aguilar}\ \emph {et~al.}(2013)\citenamefont
  {Aguilar}, \citenamefont {Ibanez},\ and\ \citenamefont
  {Papavassiliou}}]{Aguilar:2013xqa}%
  \BibitemOpen
  \bibfield  {author} {\bibinfo {author} {\bibfnamefont {A.~C.}\ \bibnamefont
  {Aguilar}}, \bibinfo {author} {\bibfnamefont {D.}~\bibnamefont {Ibanez}}, \
  and\ \bibinfo {author} {\bibfnamefont {J.}~\bibnamefont {Papavassiliou}},\
  }\href {\doibase 10.1103/PhysRevD.87.114020} {\bibfield  {journal} {\bibinfo
  {journal} {Phys. Rev.}\ }\textbf {\bibinfo {volume} {D87}},\ \bibinfo {pages}
  {114020} (\bibinfo {year} {2013})},\ \Eprint {http://arxiv.org/abs/1303.3609}
  {arXiv:1303.3609 [hep-ph]} \BibitemShut {NoStop}%
\bibitem [{\citenamefont {Huber}\ and\ \citenamefont {von
  Smekal}(2013)}]{Huber:2012kd}%
  \BibitemOpen
  \bibfield  {author} {\bibinfo {author} {\bibfnamefont {M.~Q.}\ \bibnamefont
  {Huber}}\ and\ \bibinfo {author} {\bibfnamefont {L.}~\bibnamefont {von
  Smekal}},\ }\href {\doibase 10.1007/JHEP04(2013)149} {\bibfield  {journal}
  {\bibinfo  {journal} {JHEP}\ }\textbf {\bibinfo {volume} {04}},\ \bibinfo
  {pages} {149} (\bibinfo {year} {2013})},\ \Eprint
  {http://arxiv.org/abs/1211.6092} {arXiv:1211.6092 [hep-th]} \BibitemShut
  {NoStop}%
\bibitem [{\citenamefont {Piguet}\ and\ \citenamefont
  {Sorella}(1995)}]{Piguet:1995er}%
  \BibitemOpen
  \bibfield  {author} {\bibinfo {author} {\bibfnamefont {O.}~\bibnamefont
  {Piguet}}\ and\ \bibinfo {author} {\bibfnamefont {S.~P.}\ \bibnamefont
  {Sorella}},\ }\href {\doibase 10.1007/978-3-540-49192-7} {\bibfield
  {journal} {\bibinfo  {journal} {Lect. Notes Phys. Monogr.}\ }\textbf
  {\bibinfo {volume} {28}},\ \bibinfo {pages} {1} (\bibinfo {year}
  {1995})}\BibitemShut {NoStop}%
\end{thebibliography}%

\end{document}